\documentclass[pdflatex,sn-mathphys-num]{sn-jnl}

\usepackage{graphicx}%
\usepackage{multirow}%
\usepackage{amsmath,amssymb,amsfonts}%
\usepackage{amsthm}%
\usepackage{mathrsfs}%
\usepackage[title]{appendix}%
\usepackage{xcolor}%
\usepackage{textcomp}%
\usepackage{manyfoot}%
\usepackage{booktabs}%
\usepackage{algorithm}%
\usepackage{algorithmicx}%
\usepackage{algpseudocode}%
\usepackage{listings}%
\usepackage{soul} %
\usepackage{comment}
\usepackage{float}

\theoremstyle{thmstyleone}%
\theoremstyle{thmstyletwo}%

\theoremstyle{thmstylethree}%

\begin{document}

\title[Article Title]{Substantial Diel Changes of Cloud Adjustments to Aerosols in Ship-tracks}


\author*[1,2]{\fnm{Tianle} \sur{Yuan}}\email{tianle.yuan@nasa.gov}

\author[1,3]{\fnm{Hua} \sur{Song}}

\author[4]{\fnm{Robert} \sur{Wood}}

\author[1]{\fnm{Lazaros} \sur{Oreopoulos}}

\author[1]{\fnm{Kerry} \sur{Meyer}}

\author[5]{\fnm{William} \sur{Smith}}

\author[4]{\fnm{Ryan} \sur{Eastman}}

\affil*[1]{\orgdiv{Earth Sciences Division}, \orgname{NASA Goddard Space Flight Center}, \orgaddress{ \city{Greenbelt}, \postcode{20771}, \state{State}, \country{Country}}}

\affil[2]{\orgdiv{GESTAR-II}, \orgname{University of Maryland, Baltimore County}, \orgaddress{ \city{Baltimore}, \postcode{21228}, \state{MD}, \country{USA}}}

\affil[3]{\orgname{SSAI Inc.}, \orgaddress{\city{Lanham}, \postcode{20706}, \state{MD}, \country{USA}}}

\affil[4]{\orgdiv{Department of Atmospheric Sciences},\orgname{University of Washington}, \orgaddress{\city{Seattle}, \postcode{98195}, \state{WA}, \country{USA}}}

\affil[5]{\orgdiv{Science Directorate},\orgname{NASA Langley Research Center}, \orgaddress{\city{Hampton}, \postcode{23681}, \state{VA}, \country{USA}}}



\abstract{ Human-induced changes in atmospheric aerosols have introduced a climate forcing by modifying cloud droplet number concentration, liquid water, and cloud fraction \cite{twomey_influence_1977, albrecht_aerosols_1989}. This forcing is subject to large uncertainties as cloud adjustments have not only complex dependence on background conditions \cite{forster_earths_2021,bellouin_bounding_2020}, but also temporal fluctuations, especially those driven by diel variations in solar heating \cite{wang_modelling_2010}. However, direct observations of such diel changes are still limited. Here, we present observational evidence of substantial diel changes in the cloud adjustments to aerosols within ship-tracks—linear lines of polluted clouds captured in satellite images. We developed a novel method to automatically determine the age of each ship-track segment and analyze cloud adjustments to aerosols. We show that more aged polluted clouds with extended nighttime exposure exhibit higher increases in cloud fraction. By contrast, liquid water path adjustments follow a non-monotonic pattern: they generally decrease with time before reversing trend in clouds formed at nighttime. Most of these diel contrasts are statistically significant and likely stem from differences in solar heating and cloud precipitation. The increase in cloud fraction adjustment suggests a larger aerosol effective radiative forcing, $-0.1 \sim -0.4 Wm^{-2}$, than the estimate without considering temporal variations, while the temporal changes in liquid water path adjustments may partially offset it. These findings underscore the importance of diel variations in aerosol–cloud interactions. Our approach demonstrates that ship-tracks, despite appearing as instantaneous observations, yield valuable insights into the temporal evolution of cloud adjustments.}




\maketitle

\section{Introduction}\label{sec1}
Aerosol particles can change the energy balance of the Earth’s atmosphere indirectly through interacting with clouds and modifying their brightness and coverage. Such aerosol-induced energy perturbations are known as the aerosol effective radiative forcing, including both the Twomey effect \cite{twomey_influence_1977} and cloud adjustments\cite{forster_earths_2021}. It has partially masked the global warming caused by greenhouse gases. Current best estimate of total aerosol effective radiative forcing is between -1.7 and -0.3 Wm\textsuperscript{-2} , and remains the largest source of uncertainty in our forcing estimates \cite{bellouin_bounding_2020}. The uncertainty range – 1.4Wm\textsuperscript{-2} – is nearly half of forcing from increasing greenhouse gases, which suggests that observed warming can be a result of either small or large net positive climate forcing.

Aerosols affect low cloud properties by acting as cloud condensation nuclei (CCN). Increasing CCN can lead to various changes in low cloud  droplet number concentration ($N_d$) \cite{twomey_influence_1977}, liquid water($L$ ) \cite{ackerman_impact_2004}, and coverage ($CF$)\cite{albrecht_aerosols_1989}. If $L$ remains constant, clouds become brighter with increasing CCN because of increased total droplet surface area. But aerosols can also change $L$ through modifying cloud precipitation and entrainment. Smaller droplets in more polluted clouds decrease precipitation efficiency, but increase $L$ because of that. On the other hand, smaller droplets and decreased precipitation efficiency can also enhance entrainment of dry and warm air above cloud top, which reduces $L$. Furthermore, changes in $L$ and precipitation can affect $CF$. Adjustments of $CF$ and $L$ account for a substantial portion of the total aerosol effective radiative forcing \cite{yuan_observational_2023,manshausen_invisible_2022,chen_substantial_2024}. Because current climate models cannot resolve these complex interactions, they have to use parameterization, which introduces uncertainty in AIF \cite{bellouin_bounding_2020}.

Opportunistic experiments have been used to constrain aerosol-cloud interactions with observations because they allow both quantification and attribution of cloud adjustments to aerosols while minimizing confounding factors\cite{christensen_opportunistic_2022}. Ship-tracks are such opportunistic experiments. They are quasi-linear lines that appear brighter than surrounding clouds and are formed due to ship-emitted aerosols affecting low clouds  \cite{coakley_effect_1987, hobbs_emissions_2000, wang_modelling_2010}. They have been widely studied to obtain insights into aerosol-cloud interactions\cite{bretherton_editorial_2000,christensen_microphysical_2011,toll_weak_2019,gryspeerdt_observing_2020,yuan_global_2022}. For example, Toll et al. \cite{toll_weak_2019} show that $L$ adjustment in global models are too positive and strong, which is independently verified by another ship-track study \cite{yuan_observational_2023}, providing important constraints for global models; ship-tracks studies show cloud adjustments depend on the background cloud state and environmental conditions\cite{christensen_microphysical_2011,gryspeerdt_observing_2020,yuan_observational_2023}. 

Ship-track studies are usually based on snapshot observations by polar orbiting satellites that usually pass over clouds at fixed time of the day. Such observations are thought to be unable to capture temporal evolution of  $CF$ and $L$ adjustments to aerosols , especially due to diel changes in solar heating\cite{wang_modelling_2010}. Strong diel changes of cloud adjustments would imply observational constraints based on ship-tracks sampled by snapthots may be biased. For example, since $L$ adjustment under nocturnal conditions in a model strongly decreases with time due to enhanced entrainment, Glassmeier et al. \cite{glassmeier_aerosol-cloud-climate_2021} suggest day-time ship-tracks studies may underestimate the negative $L$ adjustment. Other modeling studies also show that both $L$ and $CF$ adjustments change with time, but their temporal evolutions are non-monotonic for the full diel cycle \cite{wang_modeling_2009,chun_microphysical_2023,kurowski_diurnal_2025}. Observational studies of temporal evolutions of cloud adjustments in opportunistic experiments are still relatively rare. A case study shows $L$ adjustment becomes less negative in the afternoon compared to morning \cite{christensen_morning--afternoon_2009}, which is supported by analysis of large ship-tracks and virtual ship-tracks \cite{gryspeerdt_observing_2020, manshausen_invisible_2022, yuan_analyses_2025}.  Evidence of the strong decrease of $L$ with time is not found in long-lasting pollution tracks \cite{rahu_diurnal_2022}. 

Strong temporal evolutions of cloud adjustments to aerosols have important implications for aerosol-cloud interaction studies and how to improve observational constraints from opportunistic experiments. In this study, we show substantial diel changes in $CF$ and $L$ adjustments and assess their impact on estimated aerosol effective radiative forcing. We develop novel methods to determine the ages of ship-track segments and analyze cloud adjustments as a function of time since ship-emitted aerosols are first introduced into ship-track segments. Analysis of age-tagged ship-track segments reveal strong $CF$ adjustment diel changes  while temporal evolutions of $L$ adjustment is non-monotonic and their impact on forcing more muted. 

\section{Determining the age of a ship-track segment}\label{sec2}

Our study region is the Northeast Pacific  ($25^o - 60^o N, 110^o-165 ^o$W) within 250 miles of the coast between 2009 and 2021. We develop algorithms to automatically determine the age of each qualified ship-track segment \cite{yuan_analyses_2025}. First, we produce virtual ship-tracks by advecting ship-emissions for every large ship in the area that sailed within 24 hours before the Aqua Moderate Resolution Imaging Spectroradiometer (MODIS) overpass time \cite{gryspeerdt_observing_2020,manshausen_invisible_2022, yuan_observational_2023}. The ship emission time and location are based on high-resolution Automatic Identification System (AIS) data. We drive the HYSPLIT model with reanalysis meteorology fields to produce virtual ship-tracks.  Virtual ship-tracks are then matched with detected ship-tracks observed by Aqua MODIS \cite{yuan_global_2022}. Each matched ship-track is divided into 30-km long segments \cite{coakley_limits_2002}, whose age is the time difference between its emission time and MODIS overpass time. At the time of MODIS overpass, we can thus differentiate ship-track segments that are formed during the nighttime from those only experience daytime conditions. For each segment, our algorithm automatically obtains $L$, $N_d$, $CF$ and other variables for both the background and polluted clouds to derive cloud adjustments \cite{yuan_observational_2023} (Figure 1). We refer to the Method section for more details. 
\begin{figure}[H]
    \centering
    \includegraphics[width=1\linewidth]{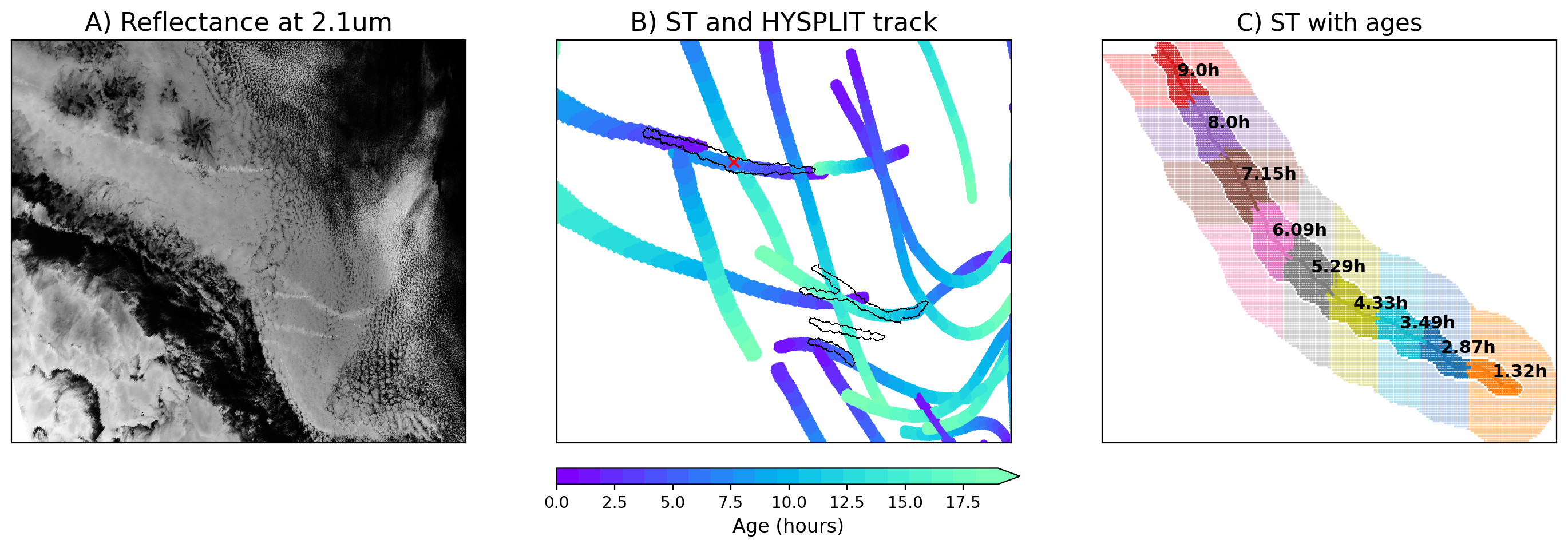}
    \caption{A) example scene of MODIS 2.1$\mu m$ reflectance image where ship-tracks can be seen. B) detected ship-tracks are plotted in black lines with one example track marked with read 'X'. C) mask of the detected ship-track and 30-pixel long segments that make up the track. The age of each segment is plotted. The ship-track and control pixels on both sides are plotted in this figure. }
\end{figure}
The examples in Figure 1 illustrate how our method works. Figure 1A shows a MODIS reflectance image. The detected ship-tracks in Figure 1A are shown as shapes enclosed by black lines in Figure 1B. Figure 1B also shows virtual ship-tracks and their ages coded in color. Detected tracks have good matches with some of the virtual tracks. The age of each segment of a matched ship-track is the same as the age of corresponding virtual tracks. The control and polluted cloud pixels for each segment are shown in Figure 1C for one matched track, and they are used to calculate cloud adjustments inside a segment. It is important to note that many virtual ship-tracks lack corresponding detected tracks for a few possible causes. For example, only small fraction of ships produce detectable ship-tracks \cite{coakley_appearance_2000,manshausen_invisible_2022,gryspeerdt_observing_2020,yuan_global_2022}; Detected ship-tracks may have no corresponding virtual tracks due to errors in reanalysis wind, or their age is longer than one day \cite{chun_microphysical_2023, toll_strong_2023}, our time limit for simulated tracks; background cloud conditions may not be favorable for ship-track detection due to overlapping clouds and cloud type \cite{yuan_global_2022,manshausen_invisible_2022}. 

\begin{figure}[H]
    \centering
    \includegraphics[width=0.8\linewidth]{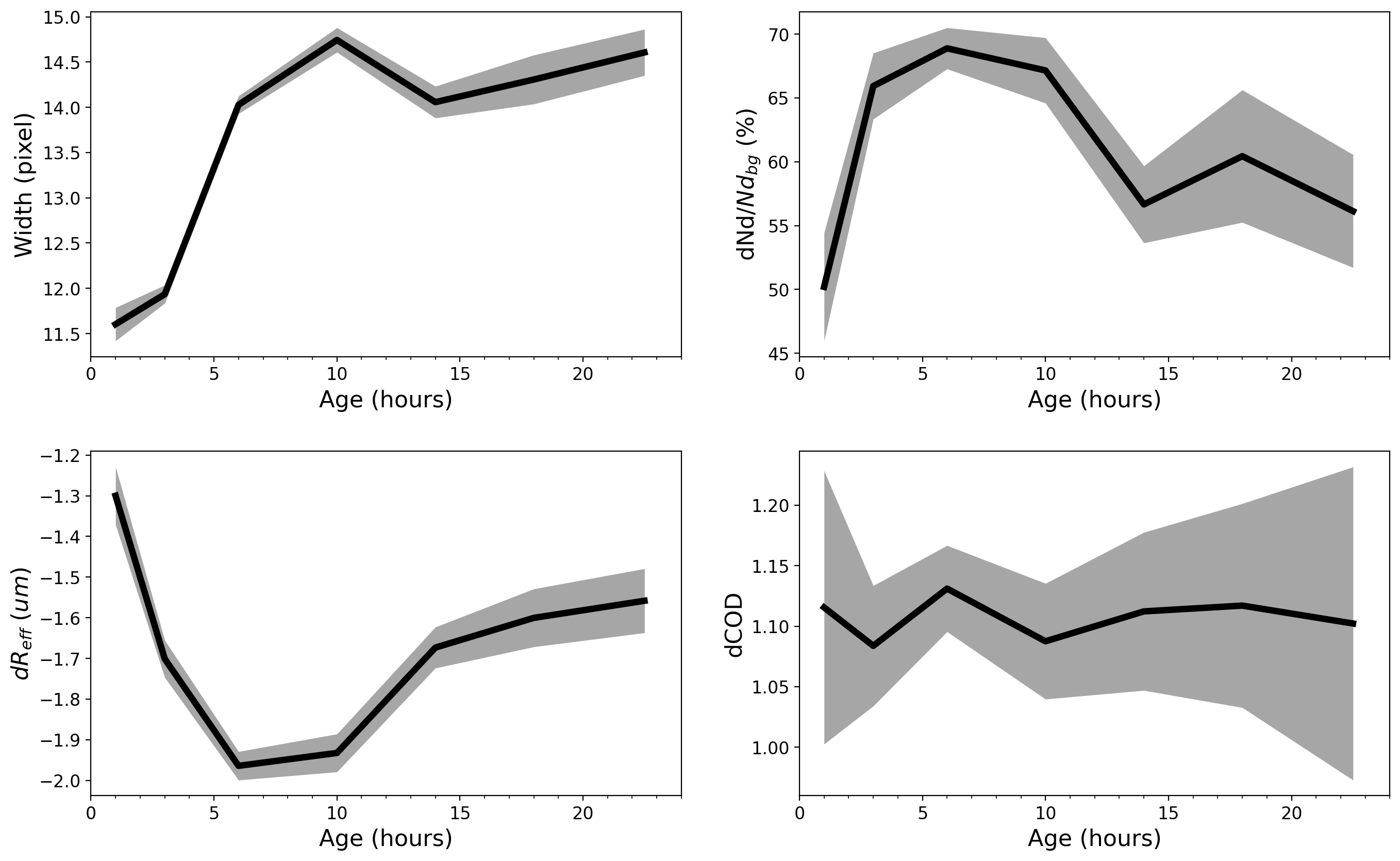}
    \caption{The temporal evolution of ship-track width, $N_d$ enhancement, changes in droplet effective radius ($\Delta{R_{eff}}$), and cloud optical depth changes ($\Delta{COD})$. Relative $\Delta N_d$ is shown here because that is the variable used to calculate aerosol effective radiative forcing and the absolute change depends on factors such as ship emission strength and background clouds.}
\end{figure}

We can study temporal evolution of ship-track properties and cloud adjustments by composite analysis of segments with different ages using the space-time interchangeability assumption (see the Method section and \cite{lensky_time-space_2006}). Temporal evolutions of ship-track width and $N_d$ enhancement ($\Delta{N_d}/N_d$) obtained using our method closely resemble those reported in the literature based on individual long tracks (Figure 2a\& b) \cite{rahu_diurnal_2022,gryspeerdt_observing_2020,manshausen_invisible_2022}, which indirectly validates the space-time interchangeability assumption. Ship-track width, defined as the ratio of area over length of a segment, expands with time rapidly at first, reaching more than 10 pixels within a few hours. Then it reaches peak value of 14 to 15 pixels by 10 hours, and plateaus afterwards, which agrees with previous results both in terms of timing and peak width\cite{durkee_composite_2000,gryspeerdt_observing_2020}. The fast horizontal spread in the initial phase could be the result of effective mixing in the boundary layer while the plateauing of the width may be related to the horizontal scale of mesoscale cellular convection that limits further effective mixing \cite{wood_spatial_2006}. We calculate $\Delta{N_d}/N_d$ based on all cloudy pixels within a ship-track segment and its average is close to 60\%, which is consistent with values reported in the literature\cite{radke_direct_1989,durkee_impact_2000,hobbs_emissions_2000,lu_marine_2007,wang_modeling_2009, christensen_microphysical_2011,russell_eastern_2013}. Our mean value of $\Delta{N_d}/N_d$ is lower than some in-situ measurements, possibly due to their limited sampling volume and a selection bias toward fresher plumes, among other factors. The temporal evolutions of $N_d$ perturbations follow similar trend of first increasing with time and then plateauing, but it decreases with age as its age further increases, especially after 10 hours. Similar temporal evolutions have been observed in individual large-tracks or invisible tracks \cite{rahu_diurnal_2022,gryspeerdt_observing_2020,manshausen_invisible_2022}. The initial increase is likely due to the time scales of gas to particle conversion such as from $SO_2$ to sulfates. The decrease of $\Delta{N_d}/N_d$ with time may be due to wet deposition by precipitation, dry deposition of particles, and diffusion. The temporal evolution of $\Delta{R_{eff}}$ mirrors that of the $\Delta{N_d}$, which is roughly consistent with the Twomey effect. The temporal evolutions of mean $\Delta {COD}$ show relatively little change, which implies competing effects of changes in $\Delta{L}$ and $\Delta{R_{eff}}$ on $COD$. 
\begin{figure} [H]
    \centering
    \includegraphics[width=0.7\linewidth]{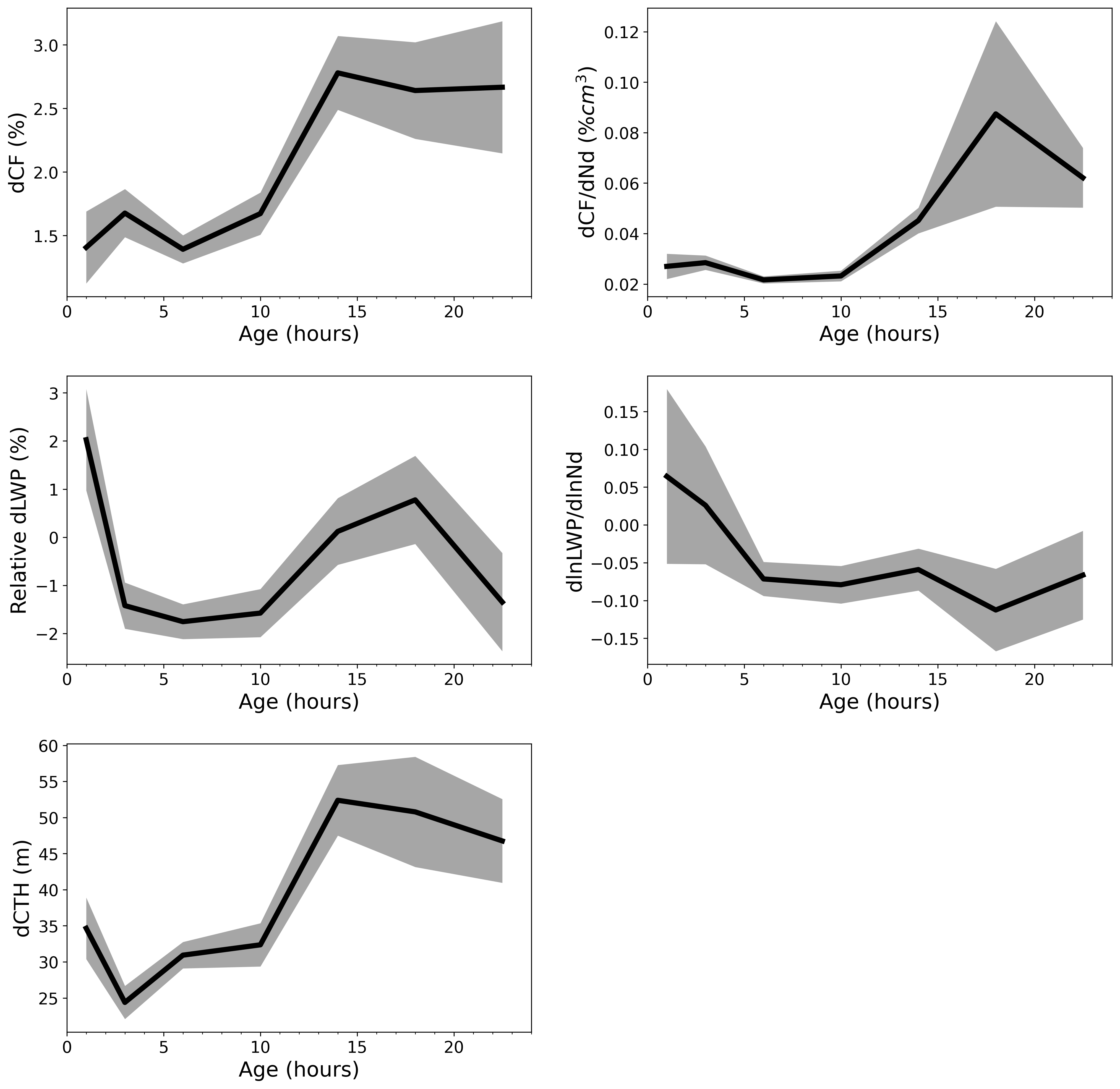}
    \caption{Temporal dynamics of cloud fraction change, cloud fraction adjustment, relative change in LWP, LWP adjustment, and changes in cloud top height. Solid black lines show the mean and shades show the standard error around the mean. }
\end{figure}

\section{Diel changes in $CF$ adjustment}\label{sec3}
While the addition of ship-emitted aerosols generally increases $CF$ in ship-tracks \cite{christensen_opportunistic_2022,yuan_observational_2023}, the $CF$ enhancement ($\Delta{CF}$) has distinct diel changes. $\Delta{CF}$ , defined as changes of $CF$ in polluted clouds inside ship-tracks relative to background clouds, shows an increase with their age starting at around 10 hr mark (Figure 3A) while average $\Delta{CF}$ is around 2\%. The 10 hour age mark roughly separates ship-track segments that are formed during nighttime conditions from those mostly only experience daytime condition for our study area because the difference between the local overpass time, \verb|~|2pm and the sunrise time,  \verb|~|6am, is 8hrs. Ship-track segments that are older than 10 hr are formed during the night and experience nighttime conditions. The $\Delta{CF}$ of ship-track clouds formed during the night is greater than those that do not experience nighttime conditions. The difference is statistically significant at above 99\% confidence level. The relative increase for nighttime clouds,  $(\Delta{CF_{night}} - \Delta{CF_{day}})/\Delta{CF_{day}}$ , is 57\%. Nighttime increase in $\Delta{CF}$ is consistent with large-eddy simulations results \cite{wang_modeling_2009,wang_modelling_2010,chun_microphysical_2023,kurowski_diurnal_2025}.

We hypothesize that the increase of $\Delta{CF}$ at night is due to drizzle suppression by aerosols. Cloud top cooling is stronger at nighttime for marine low clouds because of the absence of solar heating \cite{wood_stratocumulus_2012}. The stronger cooling tends to drive stronger convection and thicker clouds at night (see Figure 4 for an illustration), which makes them more likely to produce drizzle/precipitation during the night \cite{wang_modelling_2010}. This nighttime enhancement of drizzle has been observed for low clouds by space-borne radar measurements across low cloud areas \cite{leon_climatology_2008}. Ship-emitted aerosols therefore are thus more likely to interact with precipitating clouds and suppress drizzle at nighttime than during the day. Suppression of drizzle is more conducive to stronger $\Delta{CF}$, which explains the day-night difference in $CF$ adjustment \cite{albrecht_aerosols_1989,wang_modeling_2009,ackerman_enhancement_2003,yuan_observational_2023}.  While the diel change in $CF$ adjustment is qualitatively similar for both polluted and clean background clouds, both $\Delta{CF}$ and its diel change are larger for clean background clouds than those for more polluted background clouds, which can be explained by the effect of precipitation suppression on $CF$ of polluted clouds and the effect of precipitation on reducing $CF$ on clean clouds \cite{wang_modelling_2010, wang_manipulating_2011,wood_stratocumulus_2012}. This difference also agrees with the dependence of $CF$ adjustment inside ship-tracks on background precipitating conditions \cite{yuan_observational_2023}. This interpretation is also consistent with the decreasing trend of $\Delta{N_d}/N_d$ after the 10 hour mark because cloud precipitation is a strong sink for aerosol particles and $N_d$ and more precipitation likelihood reduces $\Delta{N_d}/N_d$.

We calculate the cloud top height difference ($\Delta{CTH}$) between ship-track and background clouds for each track segment. $\Delta{CTH}$ is significantly greater, at 99\% confidence level, for segments formed during the night than those during the day. It increases from around 30m to about 50m at night. The temporal evolution of $\Delta{CTH}$ is similar to that of $\Delta{CF}$ with an increase at around 10 hr of age that marks the nighttime formation of tracks. The increase in $\Delta{CTH}$ after 10-hour is physically consistent with drizzle suppression that tends to increase cloud top height \cite{pincus_effect_1994,wang_modelling_2010, christensen_microphysical_2012, yuan_microphysical_2011}. 

The time evolution of $CF$ adjustment to aerosols, $\frac{dCF}{dN_d}$ , has a similar increase for segments that are formed during nighttime (Figure 3c). The magnitude of relative increase in $\frac{dCF}{dN_d}$ between nighttime and daytime is stronger, \~ 300\%, than that of $\Delta{CF}$. This is because of lower $\Delta{N_d}$ for nighttime and more aged clouds while $\Delta{CF}$ is also stronger. The divergence between $\Delta{N_d}$ and $\Delta{CF}$ suggests that the stronger $\Delta{CF}$ observed in more aged and nighttime clouds reflects the cumulative impact of aerosols over time, rather than an instantaneous response \cite{chun_microphysical_2023}. 

We illustrate our hypothesis regarding the diel changes of cloud adjustments in Figure 4. In general, nighttime marine low clouds are thicker and more likely to drizzle driven by stronger cloud top radiative cooling in the absence of solar heating during the day. Ship-tracks formed during the night therefore reside in background clouds that are more likely drizzling, and precipitation suppression by ship-emitted aerosols leads to stronger $\Delta{CF}$ than those formed during daytime \cite{yuan_observational_2023,wang_modelling_2010,chun_microphysical_2023,kurowski_diurnal_2025}. Diel changes of $\Delta{N_d}$ and $\Delta{CTH}$ are also consistent with this view. The timing of the change in all these cloud variables is also consistent with each other and this interpretation. The fact that we can observe this diel change suggests that cloud adjustments last for hours, not just instantaneous, which makes them meaningful for the aerosol-cloud interaction studies \cite{chun_microphysical_2023, wang_modelling_2010,glassmeier_aerosol-cloud-climate_2021}.

\begin{figure}
    \centering
    \includegraphics[width=0.9\linewidth]{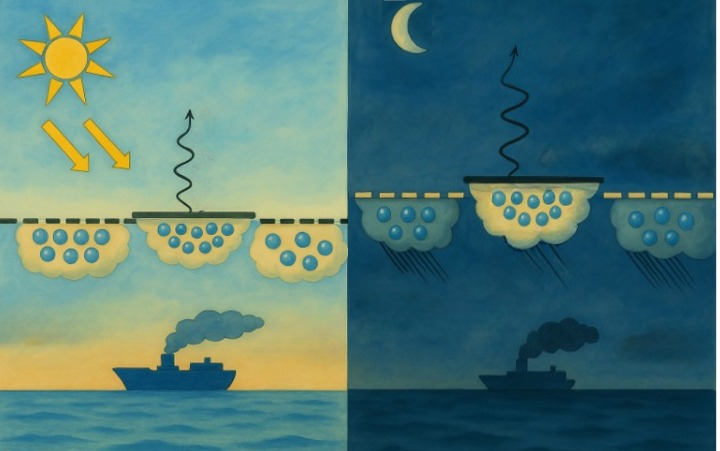}
      \caption{Schematic of the diel changes in Planetary boundary layer (PBL) clouds in general and response of clouds in ship-tracks specifically. Background PBL clouds are often thicker and more likely to precipitate at night that during the day. PBL top (white dashed line) is deeper at night  than during the day (black dashed line). Reduced precipitation efficiency inside ship-tracks tend to increase $\Delta{CF}$ and $\Delta{CTH}$.  Polluted cloud tops are outlined by solid black lines.}
\end{figure}

\section{Diel changes in $L$ adjustment}\label{sec4}
Cloud liquid water path ($L$) relative change ($\frac{\Delta{L}}{L}$) to aerosols has non-monotonic diel changes (Figure 3X). $\frac{\Delta{L}}{L}$ first increases in polluted clouds in the first couple of hours and quickly drops to negative values with time. It then continues to decrease to about 10 hr mark, after which it reverses the trend and increases with time, while remaining negative. The initial positive $\frac{\Delta{L}}{L}$ in the first hours might be an artifact because, unlike $N_d$, the timescale of $L$ adjustment cannot be instantaneous. The decrease of $\frac{\Delta{L}}{L}$ with time afterwards agrees with both modeling and observational results \cite{wang_modeling_2009,gryspeerdt_observing_2020,chun_microphysical_2023,rahu_diurnal_2022}, which indicates the effect of entrainment drying of clouds due to added aerosols and smaller droplets. Precipitation suppression is likely the main driver of the observed increasing trend in $\frac{\Delta{L}}{L}$ after 10 hr. The timing is again consistent with the precipitation suppression hypothesis during nighttime and agrees with the timing of $\Delta{CF}$ increase. 

The diel change of $L$ adjustment, calcluated as $\frac{dln\ L\ }{dln\ N_d} $, has similar evolution, except it  plateaus after 10hrs. The plateauing is an offset between a decrease in $\frac{\Delta{N_d}}{N_d}
$ and a decrease in the absolute value of $\frac{\Delta{L}}{L}$, i.e., becoming less negative, with time. Similar temporal dynamics of  $\frac{dln\ L\ }{dln\ N_d} $as ours have been found in observational and modeling studies \cite{wang_modeling_2009,rahu_diurnal_2022,chun_microphysical_2023}. The initial decreasing trend of $\frac{dln\ L\ }{dln\ N_d} $ with age is consistent with results in Glassmeier et al. \cite{glassmeier_aerosol-cloud-climate_2021}. As clouds experiencing more nighttime conditions, $L$ in polluted clouds recover more relative to the background, which stops the decreasing trend of $\frac{dln\ L\ }{dln\ N_d} $, which explains the plateauing for nighttime clouds. Quantitatively, our observed value of $\frac{dln\ L\ }{dln\ N_d} $ reaches around -0.1 at 10 hr age while it is around -0.2 at the same age \cite{glassmeier_aerosol-cloud-climate_2021}. The quantitative difference may reflect model-observation difference in a few factors. For example, our values before 10 hr of age most reflect daytime condition with solar heating while it is perpetual nocturnal in \cite{glassmeier_aerosol-cloud-climate_2021}. 

\section{Impact on forcing estimates}\label{sec5}
Younger tracks are more likely to be detected because of larger $\Delta{N_d}$ and smaller decrease in $L$. Based on the observed diel changes of cloud adjustments, this can bias estimates of cloud adjustments toward smaller $\Delta{CF}$ and possibly more negative $\Delta{L}$. We assess the impact of age selection bias in track detection on the estimate of aerosol effective radiative forcing through a simple calculation. We take the diel changes of $CF$ and $L$ adjustments as their true temporal evolutions (Figures 2 and 3). In other words, we assume each polluted cloud would have the same chance of experiencing the full diel cycle. The diel mean of cloud adjustments is therefore the mean of the diel variations in $CF$ and $L$ (Figure 3). We also calculate the simple mean of $\frac{dln\ L\ }{dln\ N_d} $ and $\frac{dCF}{dN_d}$ as the naive, but biased estimate of cloud adjustments without considering the temporal evolutions and sampling bias. The naive estimate introduces biases of 26\% and -13\% for $\frac{dCF}{dN_d}$ and $\frac{dln\ L\ }{dln\ N_d} $, respectively. In other words, the cooling effect of $CF$ adjustment is underestimated while warming impact of $L$ adjustment may be overestimated. The underestimate in $CF$ adjustment translates to additional -0.1 to -0.4 $Wm^{-2}$ of global aerosol effective radiative forcing \cite{yuan_observational_2023,yuan_abrupt_2024}, or 13\% to 50\% of the Twomey effect. Since the global mean forcing from $L$ adjustment is small in our estimate\cite{yuan_observational_2023}, the overall impact from its diel change remains small. 

\section{Discussion and conclusions}\label{sec6}

Our results highlight the substantial diel changes in cloud adjustments to aerosols and the pronounced role of day-night differences in solar heating for the diel change of cloud adjustments. This day-night contrast is an important feature of aerosol-cloud interactions that has important implications for estimating total aerosol effective radiative forcing. The exact underlying processes that are responsible for this diel change remain to be explored, but likely include interactions of multiple processes such as changes in cloud-top radiative cooling rate, entrainment and its interaction with cloud microphysics, surface heat and moisture flux, and precipitation \cite{wang_modelling_2010,wood_stratocumulus_2012,glassmeier_aerosol-cloud-climate_2021,chun_microphysical_2023,kurach_large-scale_2018}. Such diel changes in cloud adjustments may last multiple days as suggested by case studies of ship-tracks and modeling work \cite{gryspeerdt_observing_2020,rahu_diurnal_2022, chun_microphysical_2023, goren_satellite_2012}, whose cumulative impact may increase the overall impact of CF adjustments. 

Our method shows that ship and pollution tracks provide opportunities to understand the temporal dynamics of aerosol-cloud interactions and constrain modeling results despite being instantaneous measurements.  While previous studies use tracks with significant length and detectable heads to estimate segment age \cite{gryspeerdt_observing_2020,rahu_diurnal_2022,toll_strong_2023}, our method can determine the age of detected tracks without head or significant length, thereby expanding the sampling capacity and providing useful complementary information. With more access to AIS data, our approach can better study the temporal variability over more regions and under more diverse conditions.

Given the pronounced diel changes, analyses of tracks that disregard the temporal dynamics of aerosol-cloud interactions may represent a certain average of cloud responses with varying ages, in accordance with previously raised cautions \cite{glassmeier_aerosol-cloud-climate_2021}. Recent statistical analyses of observational data that are not based on opportunistic experiments also found notable diurnal changes \cite{smalley_diurnal_2024,qiu_daytime_2024}, hinting that temporal changes of aerosol-cloud interactions due to variation in solar heating may be a general feature. We suggest that resolving the temporal evolutions of aerosol-cloud interactions is important to reveal new insights and provide better observational constraints of aerosol effective radiative forcing. 

Uncertainty in cloud retrievals can affect our quantitative results. A potential example is the near instantaneous increase of $L$ in response to ship pollution, which does not seem physical and requires investigations on how retrieval artifacts may play a role. However, our results rely on the relative difference between polluted and background clouds that are equally affected by retrieval biases. We thus do not expect our results to be less systematically impacted.

Trajectory modeling is an excellent tool, but factors such as quality of reanalysis winds introduce errors in the locations of virtual ship-tracks, which limits our sample size because of unmatched ship-tracks. This impact on sample selection becomes larger with increasing length of simulation periods, which limits our ability to study temporal evolutions at longer time scales. We did not use all virtual ship tracks in our study due to this concern as positional errors in virtual tracks could have systematic biases or potentially introduce artificial signals that otherwise are not present. Further development that addresses this issue could make virtual ship-tracks more useful.

\section{Methods}\label{sec11}

\textbf{Space-time interchangeability}

We adopt the space-time interchangeability \cite{rosenfeld_satellite-based_1998, lensky_time-space_2006} assumption to study the temporal evolution of cloud adjustments in ship-tracks. With this assumption, we assume that ship-track segments with various ages but from different ship-tracks can reliably provide a good approximation of the typical temporal evolution of a ship-track. In other words, ensemble of segments that have different ages from different locations allow us to study the temporal evolution of a ship-track. This interchangeability assumption is not perfect, but provides useful approximation of the composite behavior of temporal dynamics of cloud adjustments to aerosols. The temporal evolutions of several ship-track properties such as $R_{eff}$, width and $\Delta N_d$ obtained with this approach agree well with results in the literature that track temporal evolution in individual long ship-tracks, as shown in the main text. Our approach takes advantage of the numerous detected ship-tracks and allow us to expand the scale of our analysis, which helps to build our results on more samples. It is relatively straightforward to determine the ages of ship-track segments if a head of ship-track can be clearly determined and the ship's speed is known. However, it is not the case for vast majority of detected ship-tracks. Instead, in our study, the age of a ship-track segment is determined by that of a matched virtual ship-track for which we have precise knowledge of its age, avoiding the presence of a ship-track head.

\textbf{Study region and time}

This study is conducted over the Northeast Pacific between 2009 and 2020. The geographic region is between (25 $^oN$, 165 $^oW$) and ( 60 $^oN$, 110 $^oW$) within around 250 miles of the coast. We choose this area because of the availability of automated identification system (AIS) data and its coverage of stratocumulus decks off the coast of California and low clouds in the mid-latitudes associated with cyclones.

\textbf{Data}

The AIS data have native temporal resolution of one minute, but are subsampled to be hourly data because the reanalysis data do not have matching high time-resolution. We apply a few filters on the ship AIS data. Based on the length, width, draft and speed of individual ships, we ignore ships whose $speed^2 \cdot width \cdot length <= 5\times10^5 \space m^4s^{-2}$ or $speed^2 \cdot width \cdot draft <= 5 \times 10^4 \space m^4s^{-2}$ . This is done to reduce the amount of calculation needed to run the trajectory model for each ship since smaller ships are unlikely to produce a ship-track. After the filtering, we are left with more than 6million ship samples. We ran the HYSPLIT model for these ships, each for 30 hours before local overpass time \cite{stein_noaas_2015}.  323,253 of them are within Aqua MODIS granules for matching up with detected ship-tracks. MERRA-2 reanalysis meteorology fields \cite{gelaro_modern-era_2017}are used to advect virtual emissions by these ships to produce simulated ship-tracks using the HYSPLIT model. The MERRA-2 data are 3-hourly at a resolution of 0.5$^o$ X 0.625$^o$. We compare results from MERRA-2 and ERA-5 and find their performance is very similar.  Aqua MODIS collection 6.1 retrievals \cite{platnick_modis_2017} are used in this study. We only include cloud properties retrievals of liquid cloud pixels. 

\textbf{Virtual ship-tracks and ages of ship-track segments}

The details of the method can be found in \citet{yuan_analyses_2025} and here we briefly recap the procedure. AIS data are first sub-sampled from native resolutions to hourly resolution. At each hour, we release an air parcel at the height of 950 hPa and use the HYSPLIT model to simulate its forward trajectory until the time of MODIS overpass. Each day, we simulate ships that within our study region and within 24 hours of MODIS overpassing time. We choose a cut-off of 24 hours because errors in reanalysis winds and trajectory modeling grow with time and we want to avoid simulated tracks suffer from excessive position errors while 24hours give us the full diel cycle. Once we obtain the simulated ship-tracks of each ship, we collocate them with available MODIS granules and keep only those that intersect with these granules that day. 

For each MODIS granule that has detected ship-tracks, we iterate through all detected ship-tracks and match them with simulated ship-tracks. We test a few conditions for matching, such as Frechet distance between the detected and simulated ship-tracks and maximum overlapping pixels between the two. We find the latter to be more versatile because both detected and simulated ship-tracks have intersect with each other, creating complex shapes that are difficult for Frechet distance method to be effective since we do not have the true order information on the detected ship-tracks. The age of each segment of a detected ship-track is determined to be that of the matching segment of the virtual ship-track \cite{yuan_analyses_2025}. The age of the virtual ship-track segment is the time difference between the release of virtual emission parcel and MODIS observation time. 

\textbf{Analyze cloud adjustments}

Similar to Yuan et al. \cite{yuan_observational_2023}, we automatically find background pixels for each ship-track segment. An example is shown in Figure 1. We consider the impact of overlap between high and low clouds on the true low cloud fraction \cite{yuan_observational_2023}. Only low cloud pixels are included in calculating cloud properties. $N_d$ is calculated using cloud retrievals of $R_{eff}$ and $COD$ \cite{grosvenor_remote_2018}. For each segment, we calculate cloud changes using the means of ship-track and background pixels.  

Distributions of $\Delta{L}$ are known to be not normal and have many outliers \cite{manshausen_invisible_2022}.  $\Delta{N_d}$ can be small in some of the segments. When calculating $\frac{dln\ L\ }{dln\ N_d} $, the combination of large $\Delta{L}$ and small $\Delta{N_d}$ can introduce outlier values. We try to remove top 5\% absolute values from calculated $\frac{dln\ L\ }{dln\ N_d} $ or use kernel density estimation method to remove outliers before calculate the mean $\frac{dln\ L\ }{dln\ N_d} $. We also try to adopt the method used in previous studies (e.g., \cite{manshausen_invisible_2022}) where we calculate mean $\Delta{L}$ and $\Delta{N_d}$ for a set of samples and then determine $\frac{dln\ {L}\ }{dln\ N_d} \approx \frac{dln\ \overline{L}\ }{dln\ \overline{N_d}}$. Averaging $\Delta{L}$  first makes it less prone to samples that have anomalously large values of $\frac{ \Delta{L}\ }{\Delta N_d}$.  

\textbf{Data availability}

MODIS data are freely available here: https://www.earthdata.nasa.gov/centers/laads-daac. MERRA-2 reanalysis data are also freely availalbe at https://gmao.gsfc.nasa.gov/reanalysis/merra-2/. The HYSPLIT model is available at https://www.ready.noaa.gov/HYSPLIT.php. 

\backmatter

\bmhead{Supplementary information}

\renewcommand{\thefigure}{S\arabic{figure}} 
\setcounter{figure}{0} 

\begin{figure}[H]
    \centering
    \includegraphics[width=1\linewidth]{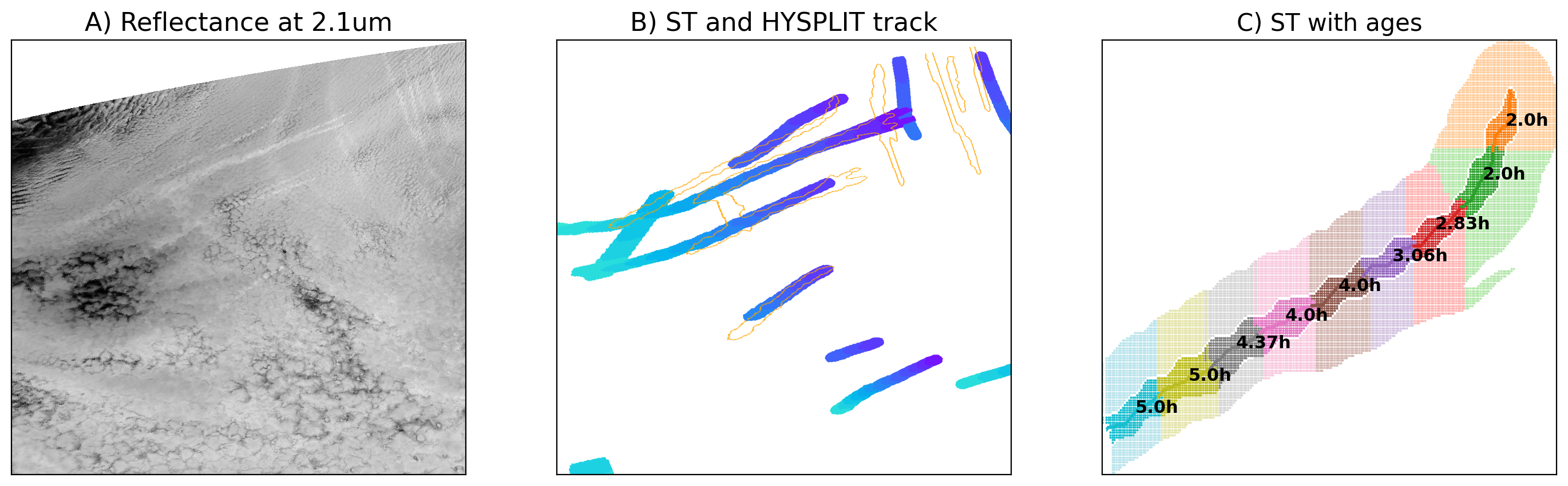}
      \caption{Similar to Figure 1, but for another time. The example is taken on Aug 19, 2016 at 2145 UTC time. }
\end{figure}

\begin{figure}[H]
    \centering
    \includegraphics[width=1\linewidth]{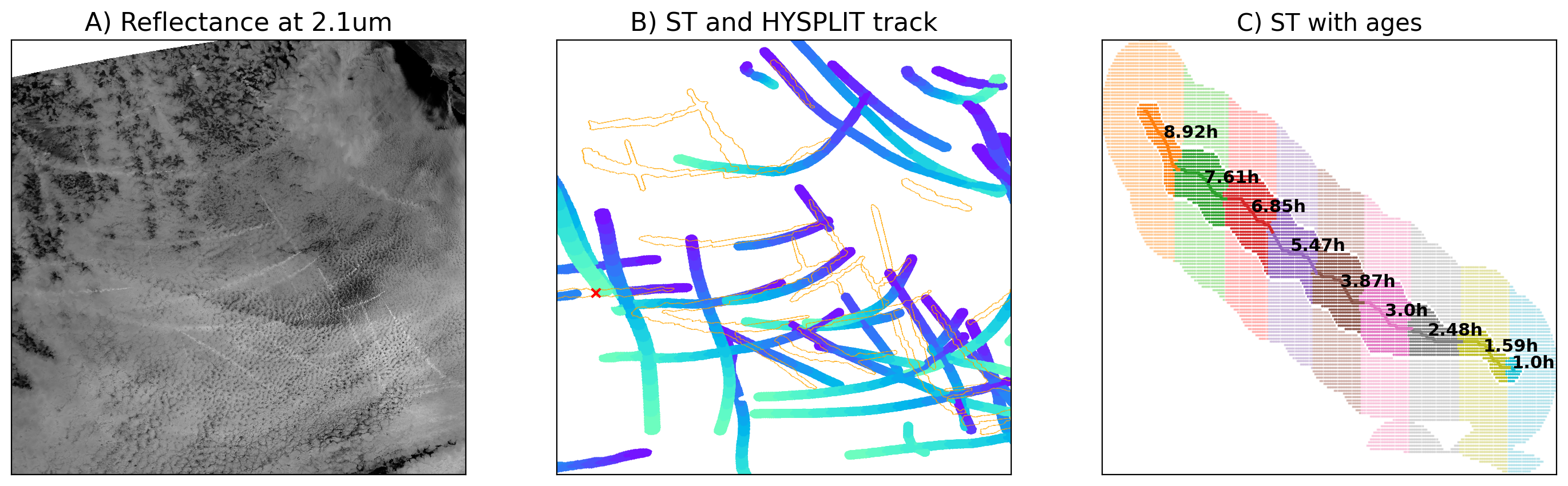}
      \caption{Similar to Figure 1, but for another time. The example is taken on Aug 2, 2018 at 2140 UTC time. }
\end{figure}

\begin{figure}[H]
    \centering
    \includegraphics[width=1\linewidth]{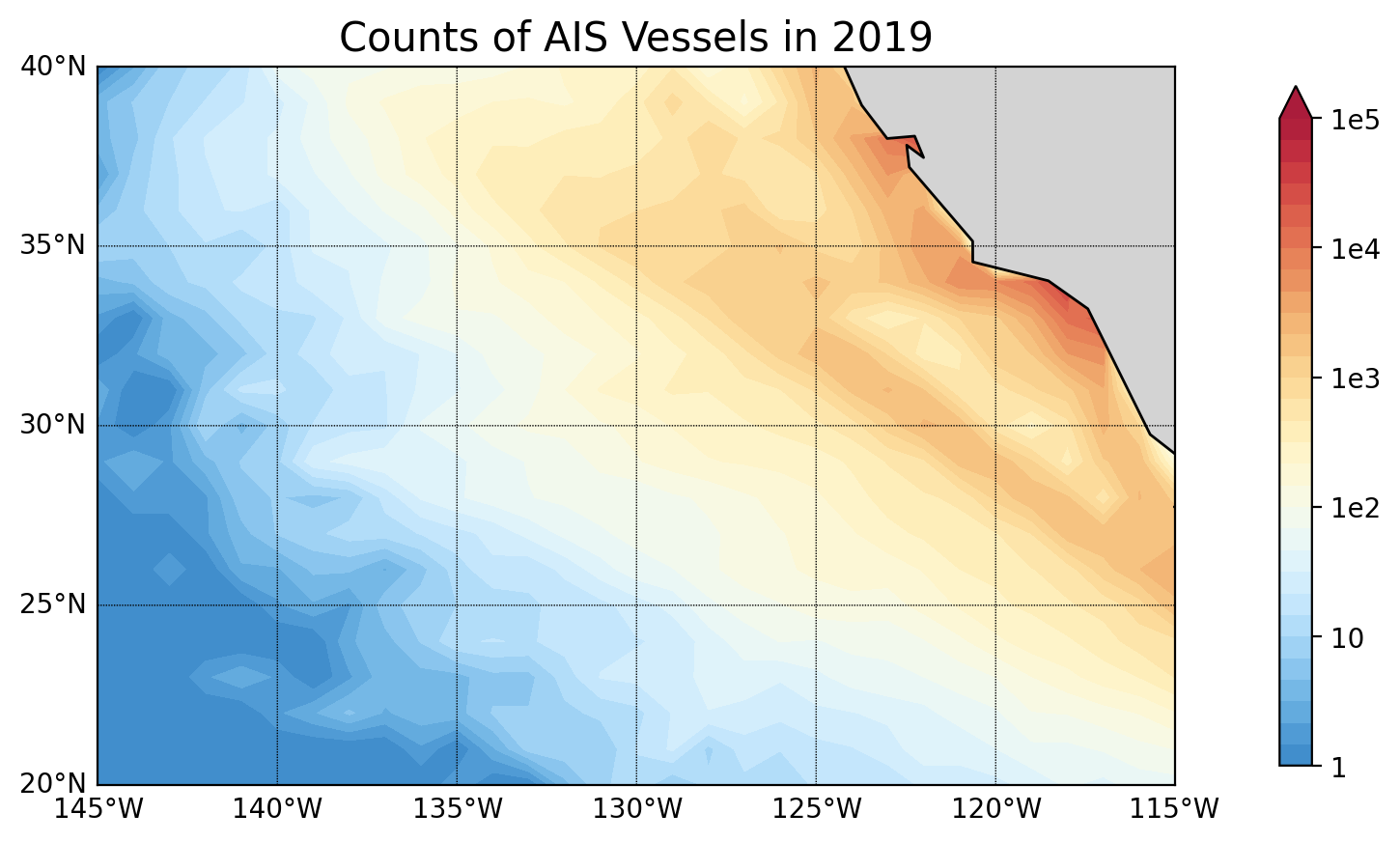}
      \caption{Number of ship occurrences for 2019 over the ocean off coast of California using AIS data. }
\end{figure}

\begin{figure}[h]
    \centering
    \includegraphics[width=1\linewidth]{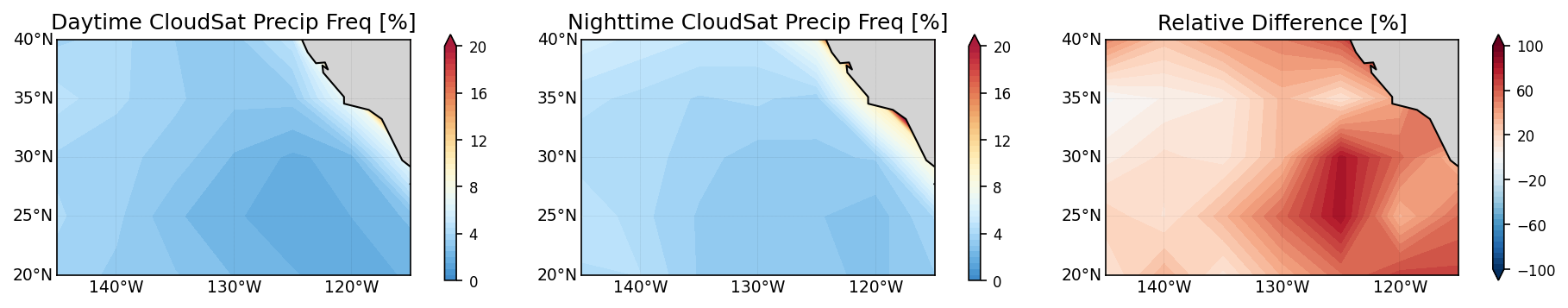}
      \caption{Low cloud precipitation frequency during daytime (a), nighttime (b), and their relative difference (c). The precipitation frequency is calculated as the ratio  between CloudSat low cloud precipitation occurrence and the CALIOP low cloud occurrence. }
\end{figure}

\begin{figure}[h]
    \centering
    \includegraphics[width=1\linewidth]{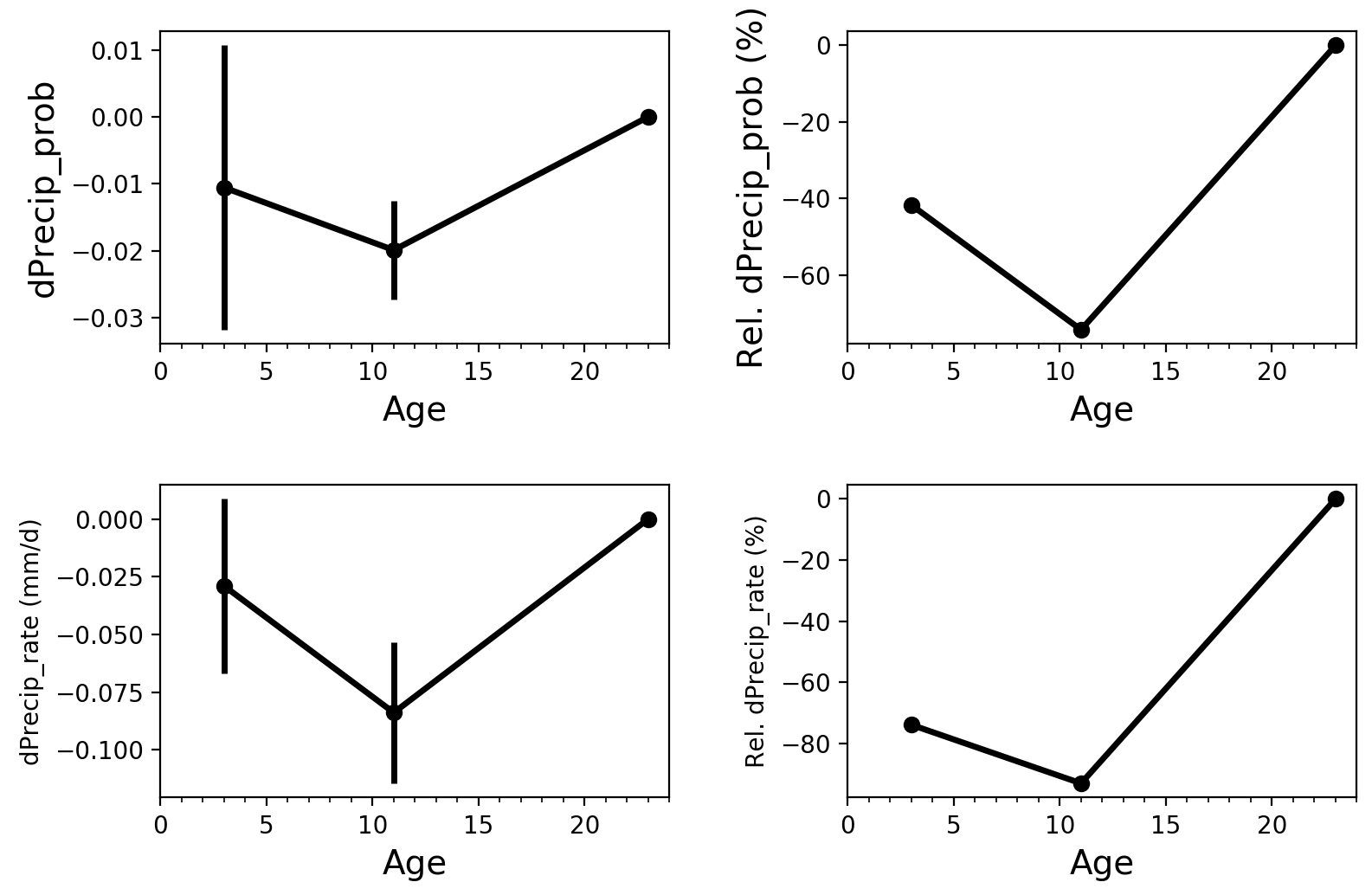}
      \caption{Temporal changees in responses of precipitation absolute and relative changes of frequency and intensity. Due to small collocated sample sizes, we divide the data into three bins. Precipitation is initially strongly suppressed, but those samples that experience more nighttime conditions and more aged show muted difference.}
\end{figure}
\bmhead{Acknowledgements}

We acknowledge funding support from the NASA MEaSUREs and TASNPP programs (grant numbers 80NSSC24K0458 and 80NSSC24M0045), the NOAA ERB program (grant NA23OAR4310299 ), and the DOE ASR program (grant DE-SC0024078). 

\section*{Declarations}
\begin{itemize}
\item Data availability 
MODIS, MERRA-2, and Cloudsat-CALIPSO data are available online. HYSPLIT model is open access and maintained by NOAA's Air Resource Laboratory. 
\item Author contribution
TY conceived the idea, analyzed the data, and wrote the manuscript draft. HS processed the data and made the plots. all other co-authors contributed to the writing of the manuscript. 

\end{itemize}

\bigskip


\bibliography{zotero}


\begin{thebibliography}{51}
\ifx \bisbn   \undefined \def \bisbn  #1{ISBN #1}\fi
\ifx \binits  \undefined \def \binits#1{#1}\fi
\ifx \bauthor  \undefined \def \bauthor#1{#1}\fi
\ifx \batitle  \undefined \def \batitle#1{#1}\fi
\ifx \bjtitle  \undefined \def \bjtitle#1{#1}\fi
\ifx \bvolume  \undefined \def \bvolume#1{\textbf{#1}}\fi
\ifx \byear  \undefined \def \byear#1{#1}\fi
\ifx \bissue  \undefined \def \bissue#1{#1}\fi
\ifx \bfpage  \undefined \def \bfpage#1{#1}\fi
\ifx \blpage  \undefined \def \blpage #1{#1}\fi
\ifx \burl  \undefined \def \burl#1{\textsf{#1}}\fi
\ifx \doiurl  \undefined \def \doiurl#1{\url{https://doi.org/#1}}\fi
\ifx \betal  \undefined \def \betal{\textit{et al.}}\fi
\ifx \binstitute  \undefined \def \binstitute#1{#1}\fi
\ifx \binstitutionaled  \undefined \def \binstitutionaled#1{#1}\fi
\ifx \bctitle  \undefined \def \bctitle#1{#1}\fi
\ifx \beditor  \undefined \def \beditor#1{#1}\fi
\ifx \bpublisher  \undefined \def \bpublisher#1{#1}\fi
\ifx \bbtitle  \undefined \def \bbtitle#1{#1}\fi
\ifx \bedition  \undefined \def \bedition#1{#1}\fi
\ifx \bseriesno  \undefined \def \bseriesno#1{#1}\fi
\ifx \blocation  \undefined \def \blocation#1{#1}\fi
\ifx \bsertitle  \undefined \def \bsertitle#1{#1}\fi
\ifx \bsnm \undefined \def \bsnm#1{#1}\fi
\ifx \bsuffix \undefined \def \bsuffix#1{#1}\fi
\ifx \bparticle \undefined \def \bparticle#1{#1}\fi
\ifx \barticle \undefined \def \barticle#1{#1}\fi
\bibcommenthead
\ifx \bconfdate \undefined \def \bconfdate #1{#1}\fi
\ifx \botherref \undefined \def \botherref #1{#1}\fi
\ifx \url \undefined \def \url#1{\textsf{#1}}\fi
\ifx \bchapter \undefined \def \bchapter#1{#1}\fi
\ifx \bbook \undefined \def \bbook#1{#1}\fi
\ifx \bcomment \undefined \def \bcomment#1{#1}\fi
\ifx \oauthor \undefined \def \oauthor#1{#1}\fi
\ifx \citeauthoryear \undefined \def \citeauthoryear#1{#1}\fi
\ifx \endbibitem  \undefined \def \endbibitem {}\fi
\ifx \bconflocation  \undefined \def \bconflocation#1{#1}\fi
\ifx \arxivurl  \undefined \def \arxivurl#1{\textsf{#1}}\fi
\csname PreBibitemsHook\endcsname

\bibitem[\protect\citeauthoryear{TWOMEY}{1977}]{twomey_influence_1977}
\begin{botherref}
\oauthor{\bsnm{TWOMEY}, \binits{S.}}:
The influence of pollution on the shortwave albedo of clouds.
Journal Of The Atmospheric Sciences
(1977)
\end{botherref}
\endbibitem

\bibitem[\protect\citeauthoryear{Albrecht}{1989}]{albrecht_aerosols_1989}
\begin{barticle}
\bauthor{\bsnm{Albrecht}, \binits{B.}}:
\batitle{Aerosols, {Cloud} {Microphysics}, and {Fractional} {Cloudiness}}.
\bjtitle{Science}
\bvolume{245}(\bissue{4923}),
\bfpage{1227}
(\byear{1989})
\doiurl{10.1126/science.245.4923.1227}
\end{barticle}
\endbibitem

\bibitem[\protect\citeauthoryear{Forster et~al.}{2021}]{forster_earths_2021}
\begin{bchapter}
\bauthor{\bsnm{Forster}, \binits{P.}},
\bauthor{\bsnm{Storelvmo}, \binits{T.}},
\bauthor{\bsnm{Armour}, \binits{K.}},
\bauthor{\bsnm{Collins}, \binits{W.}},
\bauthor{\bsnm{Dufresne}, \binits{J.-L.}},
\bauthor{\bsnm{Frame}, \binits{D.}},
\bauthor{\bsnm{Lunt}, \binits{D.J.}},
\bauthor{\bsnm{Mauritsen}, \binits{T.}},
\bauthor{\bsnm{Palmer}, \binits{M.D.}},
\bauthor{\bsnm{Watanabe}, \binits{M.}},
\bauthor{\bsnm{Wild}, \binits{M.}},
\bauthor{\bsnm{Zhang}, \binits{H.}}:
\bctitle{The {Earth}’s {Energy} {Budget}, {Climate} {Feedbacks}, and {Climate} {Sensitivity}}.
In: \bbtitle{Climate {Change} 2021: {The} {Physical} {Science} {Basis}. {Contribution} of {Working} {Group} {I} to the {Sixth} {Assessment} {Report} of the {Intergovernmental} {Panel} on {Climate} {Change}}.
\bpublisher{Cambridge University Press}, \blocation{???}
(\byear{2021})
\end{bchapter}
\endbibitem

\bibitem[\protect\citeauthoryear{Bellouin et~al.}{2020}]{bellouin_bounding_2020}
\begin{barticle}
\bauthor{\bsnm{Bellouin}, \binits{N.}},
\bauthor{\bsnm{Quaas}, \binits{J.}},
\bauthor{\bsnm{Gryspeerdt}, \binits{E.}},
\bauthor{\bsnm{Kinne}, \binits{S.}},
\bauthor{\bsnm{Stier}, \binits{P.}},
\bauthor{\bsnm{Watson-Parris}, \binits{D.}},
\bauthor{\bsnm{Boucher}, \binits{O.}},
\bauthor{\bsnm{Carslaw}, \binits{K.S.}},
\bauthor{\bsnm{Christensen}, \binits{M.}},
\bauthor{\bsnm{Daniau}, \binits{A.-L.}},
\bauthor{\bsnm{Dufresne}, \binits{J.-L.}},
\bauthor{\bsnm{Feingold}, \binits{G.}},
\bauthor{\bsnm{Fiedler}, \binits{S.}},
\bauthor{\bsnm{Forster}, \binits{P.}},
\bauthor{\bsnm{Gettelman}, \binits{A.}},
\bauthor{\bsnm{Haywood}, \binits{J.M.}},
\bauthor{\bsnm{Lohmann}, \binits{U.}},
\bauthor{\bsnm{Malavelle}, \binits{F.}},
\bauthor{\bsnm{Mauritsen}, \binits{T.}},
\bauthor{\bsnm{McCoy}, \binits{D.T.}},
\bauthor{\bsnm{Myhre}, \binits{G.}},
\bauthor{\bsnm{Mülmenstädt}, \binits{J.}},
\bauthor{\bsnm{Neubauer}, \binits{D.}},
\bauthor{\bsnm{Possner}, \binits{A.}},
\bauthor{\bsnm{Rugenstein}, \binits{M.}},
\bauthor{\bsnm{Sato}, \binits{Y.}},
\bauthor{\bsnm{Schulz}, \binits{M.}},
\bauthor{\bsnm{Schwartz}, \binits{S.E.}},
\bauthor{\bsnm{Sourdeval}, \binits{O.}},
\bauthor{\bsnm{Storelvmo}, \binits{T.}},
\bauthor{\bsnm{Toll}, \binits{V.}},
\bauthor{\bsnm{Winker}, \binits{D.}},
\bauthor{\bsnm{Stevens}, \binits{B.}}:
\batitle{Bounding {Global} {Aerosol} {Radiative} {Forcing} of {Climate} {Change}}.
\bjtitle{Reviews of Geophysics}
\bvolume{58}(\bissue{1}),
\bfpage{2019}--\blpage{000660}
(\byear{2020})
\doiurl{10.1029/2019RG000660} .
\bcomment{\_eprint: https://onlinelibrary.wiley.com/doi/pdf/10.1029/2019RG000660}.
Accessed 2023-08-18
\end{barticle}
\endbibitem

\bibitem[\protect\citeauthoryear{Wang et~al.}{2010}]{wang_modelling_2010}
\begin{barticle}
\bauthor{\bsnm{Wang}, \binits{H.}},
\bauthor{\bsnm{Feingold}, \binits{G.}},
\bauthor{\bsnm{Wood}, \binits{R.}},
\bauthor{\bsnm{Kazil}, \binits{J.}}:
\batitle{Modelling microphysical and meteorological controls on precipitation and cloud cellular structures in {Southeast} {Pacific} stratocumulus}.
\bjtitle{Atmospheric Chemistry And Physics}
\bvolume{10}(\bissue{13}),
\bfpage{6347}--\blpage{6362}
(\byear{2010})
\doiurl{10.5194/acp-10-6347-2010}
\end{barticle}
\endbibitem

\bibitem[\protect\citeauthoryear{Ackerman et~al.}{2004}]{ackerman_impact_2004}
\begin{barticle}
\bauthor{\bsnm{Ackerman}, \binits{A.S.}},
\bauthor{\bsnm{Kirkpatrick}, \binits{M.P.}},
\bauthor{\bsnm{Stevens}, \binits{D.E.}},
\bauthor{\bsnm{Toon}, \binits{O.B.}}:
\batitle{The impact of humidity above stratiform clouds on indirect aerosol climate forcing}.
\bjtitle{Nature}
\bvolume{432}(\bissue{7020}),
\bfpage{1014}--\blpage{1017}
(\byear{2004})
\doiurl{10.1038/nature03174}
\end{barticle}
\endbibitem

\bibitem[\protect\citeauthoryear{Yuan et~al.}{2023}]{yuan_observational_2023}
\begin{barticle}
\bauthor{\bsnm{Yuan}, \binits{T.}},
\bauthor{\bsnm{Song}, \binits{H.}},
\bauthor{\bsnm{Wood}, \binits{R.}},
\bauthor{\bsnm{Oreopoulos}, \binits{L.}},
\bauthor{\bsnm{Platnick}, \binits{S.}},
\bauthor{\bsnm{Wang}, \binits{C.}},
\bauthor{\bsnm{Yu}, \binits{H.}},
\bauthor{\bsnm{Meyer}, \binits{K.}},
\bauthor{\bsnm{Wilcox}, \binits{E.}}:
\batitle{Observational evidence of strong forcing from aerosol effect on low cloud coverage}.
\bjtitle{Science Advances}
\bvolume{9}(\bissue{45}),
\bfpage{7716}
(\byear{2023})
\doiurl{10.1126/sciadv.adh7716} .
\bcomment{Publisher: American Association for the Advancement of Science}.
Accessed 2024-10-02
\end{barticle}
\endbibitem

\bibitem[\protect\citeauthoryear{Manshausen et~al.}{2022}]{manshausen_invisible_2022}
\begin{barticle}
\bauthor{\bsnm{Manshausen}, \binits{P.}},
\bauthor{\bsnm{Watson-Parris}, \binits{D.}},
\bauthor{\bsnm{Christensen}, \binits{M.W.}},
\bauthor{\bsnm{Jalkanen}, \binits{J.-P.}},
\bauthor{\bsnm{Stier}, \binits{P.}}:
\batitle{Invisible ship tracks show large cloud sensitivity to aerosol}.
\bjtitle{Nature}
\bvolume{610}(\bissue{7930}),
\bfpage{101}--\blpage{106}
(\byear{2022})
\doiurl{10.1038/s41586-022-05122-0} .
\bcomment{Number: 7930 Publisher: Nature Publishing Group}.
Accessed 2023-10-06
\end{barticle}
\endbibitem

\bibitem[\protect\citeauthoryear{Chen et~al.}{2024}]{chen_substantial_2024}
\begin{barticle}
\bauthor{\bsnm{Chen}, \binits{Y.}},
\bauthor{\bsnm{Haywood}, \binits{J.}},
\bauthor{\bsnm{Wang}, \binits{Y.}},
\bauthor{\bsnm{Malavelle}, \binits{F.}},
\bauthor{\bsnm{Jordan}, \binits{G.}},
\bauthor{\bsnm{Peace}, \binits{A.}},
\bauthor{\bsnm{Partridge}, \binits{D.G.}},
\bauthor{\bsnm{Cho}, \binits{N.}},
\bauthor{\bsnm{Oreopoulos}, \binits{L.}},
\bauthor{\bsnm{Grosvenor}, \binits{D.}},
\bauthor{\bsnm{Field}, \binits{P.}},
\bauthor{\bsnm{Allan}, \binits{R.P.}},
\bauthor{\bsnm{Lohmann}, \binits{U.}}:
\batitle{Substantial cooling effect from aerosol-induced increase in tropical marine cloud cover}.
\bjtitle{Nature Geoscience}
\bvolume{17}(\bissue{5}),
\bfpage{404}--\blpage{410}
(\byear{2024})
\doiurl{10.1038/s41561-024-01427-z} .
\bcomment{Publisher: Nature Publishing Group}.
Accessed 2024-10-10
\end{barticle}
\endbibitem

\bibitem[\protect\citeauthoryear{Christensen et~al.}{2022}]{christensen_opportunistic_2022}
\begin{barticle}
\bauthor{\bsnm{Christensen}, \binits{M.W.}},
\bauthor{\bsnm{Gettelman}, \binits{A.}},
\bauthor{\bsnm{Cermak}, \binits{J.}},
\bauthor{\bsnm{Dagan}, \binits{G.}},
\bauthor{\bsnm{Diamond}, \binits{M.}},
\bauthor{\bsnm{Douglas}, \binits{A.}},
\bauthor{\bsnm{Feingold}, \binits{G.}},
\bauthor{\bsnm{Glassmeier}, \binits{F.}},
\bauthor{\bsnm{Goren}, \binits{T.}},
\bauthor{\bsnm{Grosvenor}, \binits{D.P.}},
\bauthor{\bsnm{Gryspeerdt}, \binits{E.}},
\bauthor{\bsnm{Kahn}, \binits{R.}},
\bauthor{\bsnm{Li}, \binits{Z.}},
\bauthor{\bsnm{Ma}, \binits{P.-L.}},
\bauthor{\bsnm{Malavelle}, \binits{F.}},
\bauthor{\bsnm{McCoy}, \binits{I.L.}},
\bauthor{\bsnm{McCoy}, \binits{D.T.}},
\bauthor{\bsnm{McFarquhar}, \binits{G.}},
\bauthor{\bsnm{Mülmenstädt}, \binits{J.}},
\bauthor{\bsnm{Pal}, \binits{S.}},
\bauthor{\bsnm{Possner}, \binits{A.}},
\bauthor{\bsnm{Povey}, \binits{A.}},
\bauthor{\bsnm{Quaas}, \binits{J.}},
\bauthor{\bsnm{Rosenfeld}, \binits{D.}},
\bauthor{\bsnm{Schmidt}, \binits{A.}},
\bauthor{\bsnm{Schrödner}, \binits{R.}},
\bauthor{\bsnm{Sorooshian}, \binits{A.}},
\bauthor{\bsnm{Stier}, \binits{P.}},
\bauthor{\bsnm{Toll}, \binits{V.}},
\bauthor{\bsnm{Watson-Parris}, \binits{D.}},
\bauthor{\bsnm{Wood}, \binits{R.}},
\bauthor{\bsnm{Yang}, \binits{M.}},
\bauthor{\bsnm{Yuan}, \binits{T.}}:
\batitle{Opportunistic experiments to constrain aerosol effective radiative forcing}.
\bjtitle{Atmospheric Chemistry and Physics}
\bvolume{22}(\bissue{1}),
\bfpage{641}--\blpage{674}
(\byear{2022})
\doiurl{10.5194/acp-22-641-2022} .
\bcomment{Publisher: Copernicus GmbH}.
Accessed 2022-01-22
\end{barticle}
\endbibitem

\bibitem[\protect\citeauthoryear{Coakley et~al.}{1987}]{coakley_effect_1987}
\begin{barticle}
\bauthor{\bsnm{Coakley}, \binits{J.A.}},
\bauthor{\bsnm{Bernstein}, \binits{R.L.}},
\bauthor{\bsnm{Durkee}, \binits{P.A.}}:
\batitle{Effect of {Ship}-{Stack} {Effluents} on {Cloud} {Reflectivity}}.
\bjtitle{Science}
\bvolume{237}(\bissue{4818}),
\bfpage{1020}--\blpage{1022}
(\byear{1987})
\doiurl{10.1126/science.237.4818.1020} .
Accessed 2019-04-21
\end{barticle}
\endbibitem

\bibitem[\protect\citeauthoryear{Hobbs et~al.}{2000}]{hobbs_emissions_2000}
\begin{barticle}
\bauthor{\bsnm{Hobbs}, \binits{P.V.}},
\bauthor{\bsnm{Garrett}, \binits{T.J.}},
\bauthor{\bsnm{Ferek}, \binits{R.J.}},
\bauthor{\bsnm{Strader}, \binits{S.R.}},
\bauthor{\bsnm{Hegg}, \binits{D.A.}},
\bauthor{\bsnm{Frick}, \binits{G.M.}},
\bauthor{\bsnm{Hoppel}, \binits{W.A.}},
\bauthor{\bsnm{Gasparovic}, \binits{R.F.}},
\bauthor{\bsnm{Russell}, \binits{L.M.}},
\bauthor{\bsnm{Johnson}, \binits{D.W.}},
\bauthor{\bsnm{O’Dowd}, \binits{C.}},
\bauthor{\bsnm{Durkee}, \binits{P.A.}},
\bauthor{\bsnm{Nielsen}, \binits{K.E.}},
\bauthor{\bsnm{Innis}, \binits{G.}}:
\batitle{Emissions from {Ships} with respect to {Their} {Effects} on {Clouds}}.
\bjtitle{Journal of the Atmospheric Sciences}
\bvolume{57}(\bissue{16}),
\bfpage{2570}--\blpage{2590}
(\byear{2000})
\doiurl{10.1175/1520-0469(2000)057<2570:EFSWRT>2.0.CO;2} .
\bcomment{Publisher: American Meteorological Society Section: Journal of the Atmospheric Sciences}.
Accessed 2022-05-14
\end{barticle}
\endbibitem

\bibitem[\protect\citeauthoryear{Bretherton}{2000}]{bretherton_editorial_2000}
\begin{barticle}
\bauthor{\bsnm{Bretherton}, \binits{C.}}:
\batitle{Editorial}.
\bjtitle{Journal of the Atmospheric Sciences}
\bvolume{57}(\bissue{16}),
\bfpage{2521}--\blpage{2521}
(\byear{2000})
\doiurl{10.1175/1520-0469(2000)057<2521:MASTE>2.0.CO;2} .
Accessed 2019-04-21
\end{barticle}
\endbibitem

\bibitem[\protect\citeauthoryear{Christensen and Stephens}{2011}]{christensen_microphysical_2011}
\begin{barticle}
\bauthor{\bsnm{Christensen}, \binits{M.W.}},
\bauthor{\bsnm{Stephens}, \binits{G.L.}}:
\batitle{Microphysical and macrophysical responses of marine stratocumulus polluted by underlying ships: {Evidence} of cloud deepening}.
\bjtitle{Journal of Geophysical Research}
\bvolume{116}(\bissue{D3}),
\bfpage{03201}
(\byear{2011})
\doiurl{10.1029/2010JD014638} .
Accessed 2019-04-21
\end{barticle}
\endbibitem

\bibitem[\protect\citeauthoryear{Toll et~al.}{2019}]{toll_weak_2019}
\begin{barticle}
\bauthor{\bsnm{Toll}, \binits{V.}},
\bauthor{\bsnm{Christensen}, \binits{M.}},
\bauthor{\bsnm{Quaas}, \binits{J.}},
\bauthor{\bsnm{Bellouin}, \binits{N.}}:
\batitle{Weak average liquid-cloud-water response to anthropogenic aerosols}.
\bjtitle{Nature}
\bvolume{572}(\bissue{7767}),
\bfpage{51}--\blpage{55}
(\byear{2019})
\doiurl{10.1038/s41586-019-1423-9} .
Accessed 2021-01-11
\end{barticle}
\endbibitem

\bibitem[\protect\citeauthoryear{Gryspeerdt et~al.}{2020}]{gryspeerdt_observing_2020}
\begin{botherref}
\oauthor{\bsnm{Gryspeerdt}, \binits{E.}},
\oauthor{\bsnm{Goren}, \binits{T.}},
\oauthor{\bsnm{Smith}, \binits{T.W.P.}}:
Observing the timescales of aerosol-cloud interactions in snapshot satellite images.
preprint,
Clouds and Precipitation/Remote Sensing/Troposphere/Physics (physical properties and processes)
(October 2020).
\doiurl{10.5194/acp-2020-1030} .
\url{https://acp.copernicus.org/preprints/acp-2020-1030/}
Accessed 2021-02-03
\end{botherref}
\endbibitem

\bibitem[\protect\citeauthoryear{Yuan et~al.}{2022}]{yuan_global_2022}
\begin{barticle}
\bauthor{\bsnm{Yuan}, \binits{T.}},
\bauthor{\bsnm{Song}, \binits{H.}},
\bauthor{\bsnm{Wood}, \binits{R.}},
\bauthor{\bsnm{Wang}, \binits{C.}},
\bauthor{\bsnm{Oreopoulos}, \binits{L.}},
\bauthor{\bsnm{Platnick}, \binits{S.E.}},
\bauthor{\bsnm{Hippel}, \binits{S.}},
\bauthor{\bsnm{Meyer}, \binits{K.}},
\bauthor{\bsnm{Light}, \binits{S.}},
\bauthor{\bsnm{Wilcox}, \binits{E.}}:
\batitle{Global reduction in ship-tracks from sulfur regulations for shipping fuel}.
\bjtitle{Science Advances}
\bvolume{8}(\bissue{29}),
\bfpage{7988}
(\byear{2022})
\doiurl{10.1126/sciadv.abn7988} .
\bcomment{Publisher: American Association for the Advancement of Science}.
Accessed 2024-12-19
\end{barticle}
\endbibitem

\bibitem[\protect\citeauthoryear{Glassmeier et~al.}{2021}]{glassmeier_aerosol-cloud-climate_2021}
\begin{barticle}
\bauthor{\bsnm{Glassmeier}, \binits{F.}},
\bauthor{\bsnm{Hoffmann}, \binits{F.}},
\bauthor{\bsnm{Johnson}, \binits{J.S.}},
\bauthor{\bsnm{Yamaguchi}, \binits{T.}},
\bauthor{\bsnm{Carslaw}, \binits{K.S.}},
\bauthor{\bsnm{Feingold}, \binits{G.}}:
\batitle{Aerosol-cloud-climate cooling overestimated by ship-track data}.
\bjtitle{Science}
\bvolume{371}(\bissue{6528}),
\bfpage{485}--\blpage{489}
(\byear{2021})
\doiurl{10.1126/science.abd3980} .
Accessed 2021-02-03
\end{barticle}
\endbibitem

\bibitem[\protect\citeauthoryear{Wang and Feingold}{2009}]{wang_modeling_2009}
\begin{barticle}
\bauthor{\bsnm{Wang}, \binits{H.}},
\bauthor{\bsnm{Feingold}, \binits{G.}}:
\batitle{Modeling mesoscale cellular structures and drizzle in marine stratocumulus. {Part} {I}: {Impact} of drizzle on the formation and evolution of open cells}.
\bjtitle{Journal Of The Atmospheric Sciences}
\bvolume{66}(\bissue{11}),
\bfpage{3237}--\blpage{3256}
(\byear{2009})
\doiurl{10.1175/2009JAS3022.1}
\end{barticle}
\endbibitem

\bibitem[\protect\citeauthoryear{Chun et~al.}{2023}]{chun_microphysical_2023}
\begin{barticle}
\bauthor{\bsnm{Chun}, \binits{J.-Y.}},
\bauthor{\bsnm{Wood}, \binits{R.}},
\bauthor{\bsnm{Blossey}, \binits{P.}},
\bauthor{\bsnm{Doherty}, \binits{S.J.}}:
\batitle{Microphysical, macrophysical, and radiative responses of subtropical marine clouds to aerosol injections}.
\bjtitle{ATMOSPHERIC CHEMISTRY AND PHYSICS}
\bvolume{23}(\bissue{2}),
\bfpage{1345}--\blpage{1368}
(\byear{2023})
\doiurl{10.5194/acp-23-1345-2023} .
\bcomment{Num Pages: 24 Place: Gottingen Publisher: Copernicus Gesellschaft Mbh Web of Science ID: WOS:000920063400001}.
Accessed 2024-10-04
\end{barticle}
\endbibitem

\bibitem[\protect\citeauthoryear{Kurowski et~al.}{2025}]{kurowski_diurnal_2025}
\begin{botherref}
\oauthor{\bsnm{Kurowski}, \binits{M.J.}},
\oauthor{\bsnm{Lebsock}, \binits{M.D.}},
\oauthor{\bsnm{Smalley}, \binits{K.M.}}:
The {Diurnal} {Susceptibility} of {Subtropical} {Clouds} to {Aerosols}
(2025).
\doiurl{10.5194/egusphere-2025-714} .
\url{https://egusphere.copernicus.org/preprints/2025/egusphere-2025-714/}
Accessed 2025-03-30
\end{botherref}
\endbibitem

\bibitem[\protect\citeauthoryear{Christensen et~al.}{2009}]{christensen_morning--afternoon_2009}
\begin{botherref}
\oauthor{\bsnm{Christensen}, \binits{M.W.}},
\oauthor{\bsnm{Coakley}, \binits{J.A.}},
\oauthor{\bsnm{Tahnk}, \binits{W.R.}}:
Morning-to-{Afternoon} {Evolution} of {Marine} {Stratus} {Polluted} by {Underlying} {Ships}: {Implications} for the {Relative} {Lifetimes} of {Polluted} and {Unpolluted} {Clouds}
(2009)
\doiurl{10.1175/2009JAS2951.1} .
Section: Journal of the Atmospheric Sciences.
Accessed 2024-10-05
\end{botherref}
\endbibitem

\bibitem[\protect\citeauthoryear{Yuan et~al.}{2025}]{yuan_analyses_2025}
\begin{barticle}
\bauthor{\bsnm{Yuan}, \binits{T.}},
\bauthor{\bsnm{Song}, \binits{H.}},
\bauthor{\bsnm{Oreopoulos}, \binits{L.}},
\bauthor{\bsnm{Wood}, \binits{R.}},
\bauthor{\bsnm{Meyer}, \binits{K.}},
\bauthor{\bsnm{Crawford}, \binits{A.}},
\bauthor{\bsnm{Smith}, \binits{W.}},
\bauthor{\bsnm{Eastman}, \binits{R.}}:
\batitle{Analyses of {Virtual} {Ship}‐{Tracks} {Systematically} {Underestimate} {Aerosol}‐{Cloud} {Interactions} {Signals}}.
\bjtitle{Geophysical Research Letters}
\bvolume{52}(\bissue{7}),
\bfpage{2024}--\blpage{114356}
(\byear{2025})
\doiurl{10.1029/2024GL114356} .
Accessed 2025-04-05
\end{barticle}
\endbibitem

\bibitem[\protect\citeauthoryear{Rahu et~al.}{2022}]{rahu_diurnal_2022}
\begin{barticle}
\bauthor{\bsnm{Rahu}, \binits{J.}},
\bauthor{\bsnm{Trofimov}, \binits{H.}},
\bauthor{\bsnm{Post}, \binits{P.}},
\bauthor{\bsnm{Toll}, \binits{V.}}:
\batitle{Diurnal {Evolution} of {Cloud} {Water} {Responses} to {Aerosols}}.
\bjtitle{JOURNAL OF GEOPHYSICAL RESEARCH-ATMOSPHERES}
\bvolume{127}(\bissue{10}),
\bfpage{2021}--\blpage{035091}
(\byear{2022})
\doiurl{10.1029/2021JD035091} .
\bcomment{Num Pages: 15 Place: Washington Publisher: Amer Geophysical Union Web of Science ID: WOS:000800359100001}.
Accessed 2024-10-04
\end{barticle}
\endbibitem

\bibitem[\protect\citeauthoryear{Coakley and Walsh}{2002}]{coakley_limits_2002}
\begin{barticle}
\bauthor{\bsnm{Coakley}, \binits{J.A.}},
\bauthor{\bsnm{Walsh}, \binits{C.D.}}:
\batitle{Limits to the {Aerosol} {Indirect} {Radiative} {Effect} {Derived} from {Observations} of {Ship} {Tracks}}.
\bjtitle{Journal of the Atmospheric Sciences}
\bvolume{59}(\bissue{3}),
\bfpage{668}--\blpage{680}
(\byear{2002})
\doiurl{10.1175/1520-0469(2002)059<0668:LTTAIR>2.0.CO;2} .
Accessed 2021-09-30
\end{barticle}
\endbibitem

\bibitem[\protect\citeauthoryear{Coakley et~al.}{2000}]{coakley_appearance_2000}
\begin{barticle}
\bauthor{\bsnm{Coakley}, \binits{J.A.}},
\bauthor{\bsnm{Durkee}, \binits{P.A.}},
\bauthor{\bsnm{Nielsen}, \binits{K.}},
\bauthor{\bsnm{Taylor}, \binits{J.P.}},
\bauthor{\bsnm{Platnick}, \binits{S.}},
\bauthor{\bsnm{Albrecht}, \binits{B.A.}},
\bauthor{\bsnm{Babb}, \binits{D.}},
\bauthor{\bsnm{Chang}, \binits{F.-L.}},
\bauthor{\bsnm{Tahnk}, \binits{W.R.}},
\bauthor{\bsnm{Bretherton}, \binits{C.S.}},
\bauthor{\bsnm{Hobbs}, \binits{P.V.}}:
\batitle{The {Appearance} and {Disappearance} of {Ship} {Tracks} on {Large} {Spatial} {Scales}}.
\bjtitle{Journal of the Atmospheric Sciences}
\bvolume{57}(\bissue{16}),
\bfpage{2765}--\blpage{2778}
(\byear{2000})
\doiurl{10.1175/1520-0469(2000)057<2765:TAADOS>2.0.CO;2} .
Accessed 2021-09-29
\end{barticle}
\endbibitem

\bibitem[\protect\citeauthoryear{Toll and Rahu}{2023}]{toll_strong_2023}
\begin{barticle}
\bauthor{\bsnm{Toll}, \binits{V.}},
\bauthor{\bsnm{Rahu}, \binits{J.}}:
\batitle{Strong {Anthropogenic} {Cloud} {Perturbations} {Can} {Persist} for {Multiple} {Days}}.
\bjtitle{JOURNAL OF GEOPHYSICAL RESEARCH-ATMOSPHERES}
\bvolume{128}(\bissue{9}),
\bfpage{2022}--\blpage{038146}
(\byear{2023})
\doiurl{10.1029/2022JD038146} .
\bcomment{Num Pages: 12 Place: Washington Publisher: Amer Geophysical Union Web of Science ID: WOS:001000194200016}.
Accessed 2024-10-04
\end{barticle}
\endbibitem

\bibitem[\protect\citeauthoryear{Lensky and Rosenfeld}{2006}]{lensky_time-space_2006}
\begin{barticle}
\bauthor{\bsnm{Lensky}, \binits{I.M.}},
\bauthor{\bsnm{Rosenfeld}, \binits{D.}}:
\batitle{The time-space exchangeability of satellite retrieved relations between cloud top temperature and particle effective radius}.
\bjtitle{Atmospheric Chemistry And Physics}
\bvolume{6},
\bfpage{2887}--\blpage{2894}
(\byear{2006})
\end{barticle}
\endbibitem

\bibitem[\protect\citeauthoryear{Durkee et~al.}{2000}]{durkee_composite_2000}
\begin{barticle}
\bauthor{\bsnm{Durkee}, \binits{P.A.}},
\bauthor{\bsnm{Chartier}, \binits{R.E.}},
\bauthor{\bsnm{Brown}, \binits{A.}},
\bauthor{\bsnm{Trehubenko}, \binits{E.J.}},
\bauthor{\bsnm{Rogerson}, \binits{S.D.}},
\bauthor{\bsnm{Skupniewicz}, \binits{C.}},
\bauthor{\bsnm{Nielsen}, \binits{K.E.}},
\bauthor{\bsnm{Platnick}, \binits{S.}},
\bauthor{\bsnm{King}, \binits{M.D.}}:
\batitle{Composite {Ship} {Track} {Characteristics}}.
\bjtitle{Journal Of The Atmospheric Sciences}
\bvolume{57}(\bissue{16}),
\bfpage{2542}--\blpage{2553}
(\byear{2000})
\doiurl{10.1175/1520-0469(2000)057<2542:CSTC>2.0.CO;2}
\end{barticle}
\endbibitem

\bibitem[\protect\citeauthoryear{Wood and Hartmann}{2006}]{wood_spatial_2006}
\begin{barticle}
\bauthor{\bsnm{Wood}, \binits{R.}},
\bauthor{\bsnm{Hartmann}, \binits{D.L.}}:
\batitle{Spatial variability of liquid water path in marine low cloud: {The} importance of mesoscale cellular convection}.
\bjtitle{Journal Of Climate}
\bvolume{19}(\bissue{9}),
\bfpage{1748}--\blpage{1764}
(\byear{2006})
\end{barticle}
\endbibitem

\bibitem[\protect\citeauthoryear{Radke et~al.}{1989}]{radke_direct_1989}
\begin{barticle}
\bauthor{\bsnm{Radke}, \binits{L.F.}},
\bauthor{\bsnm{Coakley}, \binits{J.A.}},
\bauthor{\bsnm{King}, \binits{M.D.}}:
\batitle{Direct and {Remote} {Sensing} {Observations} of the {Effects} of {Ships} on {Clouds}}.
\bjtitle{Science}
\bvolume{246}(\bissue{4934}),
\bfpage{1146}--\blpage{1149}
(\byear{1989})
\doiurl{10.1126/science.246.4934.1146} .
Accessed 2019-04-21
\end{barticle}
\endbibitem

\bibitem[\protect\citeauthoryear{Durkee et~al.}{2000}]{durkee_impact_2000}
\begin{barticle}
\bauthor{\bsnm{Durkee}, \binits{P.A.}},
\bauthor{\bsnm{Noone}, \binits{K.J.}},
\bauthor{\bsnm{Ferek}, \binits{R.J.}},
\bauthor{\bsnm{Johnson}, \binits{D.W.}},
\bauthor{\bsnm{Taylor}, \binits{J.P.}},
\bauthor{\bsnm{Garrett}, \binits{T.J.}},
\bauthor{\bsnm{Hobbs}, \binits{P.V.}},
\bauthor{\bsnm{Hudson}, \binits{J.G.}},
\bauthor{\bsnm{Bretherton}, \binits{C.S.}},
\bauthor{\bsnm{Innis}, \binits{G.}},
\bauthor{\bsnm{Frick}, \binits{G.M.}},
\bauthor{\bsnm{Hoppel}, \binits{W.A.}},
\bauthor{\bsnm{O’Dowd}, \binits{C.D.}},
\bauthor{\bsnm{Russell}, \binits{L.M.}},
\bauthor{\bsnm{Gasparovic}, \binits{R.}},
\bauthor{\bsnm{Nielsen}, \binits{K.E.}},
\bauthor{\bsnm{Tessmer}, \binits{S.A.}},
\bauthor{\bsnm{Öström}, \binits{E.}},
\bauthor{\bsnm{Osborne}, \binits{S.R.}},
\bauthor{\bsnm{Flagan}, \binits{R.C.}},
\bauthor{\bsnm{Seinfeld}, \binits{J.H.}},
\bauthor{\bsnm{Rand}, \binits{H.}}:
\batitle{The {Impact} of {Ship}-{Produced} {Aerosols} on the {Microstructure} and {Albedo} of {Warm} {Marine} {Stratocumulus} {Clouds}: {A} {Test} of {MAST} {Hypotheses} 1i and 1ii}.
\bjtitle{Journal of the Atmospheric Sciences}
\bvolume{57}(\bissue{16}),
\bfpage{2554}--\blpage{2569}
(\byear{2000})
\doiurl{10.1175/1520-0469(2000)057<2554:TIOSPA>2.0.CO;2} .
\bcomment{Publisher: American Meteorological Society Section: Journal of the Atmospheric Sciences}.
Accessed 2022-01-27
\end{barticle}
\endbibitem

\bibitem[\protect\citeauthoryear{Lu et~al.}{2007}]{lu_marine_2007}
\begin{botherref}
\oauthor{\bsnm{Lu}, \binits{M.-L.}},
\oauthor{\bsnm{Conant}, \binits{W.C.}},
\oauthor{\bsnm{Jonsson}, \binits{H.H.}},
\oauthor{\bsnm{Varutbangkul}, \binits{V.}},
\oauthor{\bsnm{Flagan}, \binits{R.C.}},
\oauthor{\bsnm{Seinfeld}, \binits{J.H.}}:
The {Marine} {Stratus}/{Stratocumulus} {Experiment} ({MASE}): {Aerosol}-cloud relationships in marine stratocumulus.
Journal of Geophysical Research: Atmospheres
\textbf{112}(D10)
(2007)
\doiurl{10.1029/2006JD007985} .
Accessed 2019-04-21
\end{botherref}
\endbibitem

\bibitem[\protect\citeauthoryear{Russell et~al.}{2013}]{russell_eastern_2013}
\begin{barticle}
\bauthor{\bsnm{Russell}, \binits{L.M.}},
\bauthor{\bsnm{Sorooshian}, \binits{A.}},
\bauthor{\bsnm{Seinfeld}, \binits{J.H.}},
\bauthor{\bsnm{Albrecht}, \binits{B.A.}},
\bauthor{\bsnm{Nenes}, \binits{A.}},
\bauthor{\bsnm{Ahlm}, \binits{L.}},
\bauthor{\bsnm{Chen}, \binits{Y.-C.}},
\bauthor{\bsnm{Coggon}, \binits{M.}},
\bauthor{\bsnm{Craven}, \binits{J.S.}},
\bauthor{\bsnm{Flagan}, \binits{R.C.}},
\bauthor{\bsnm{Frossard}, \binits{A.A.}},
\bauthor{\bsnm{Jonsson}, \binits{H.}},
\bauthor{\bsnm{Jung}, \binits{E.}},
\bauthor{\bsnm{Lin}, \binits{J.J.}},
\bauthor{\bsnm{Metcalf}, \binits{A.R.}},
\bauthor{\bsnm{Modini}, \binits{R.}},
\bauthor{\bsnm{Mülmenstädt}, \binits{J.}},
\bauthor{\bsnm{Roberts}, \binits{G.}},
\bauthor{\bsnm{Shingler}, \binits{T.}},
\bauthor{\bsnm{Song}, \binits{S.}},
\bauthor{\bsnm{Wang}, \binits{Z.}},
\bauthor{\bsnm{Wonaschütz}, \binits{A.}}:
\batitle{Eastern {Pacific} {Emitted} {Aerosol} {Cloud} {Experiment}}.
\bjtitle{Bulletin of the American Meteorological Society}
\bvolume{94}(\bissue{5}),
\bfpage{709}--\blpage{729}
(\byear{2013})
\doiurl{10.1175/BAMS-D-12-00015.1} .
Accessed 2019-04-21
\end{barticle}
\endbibitem

\bibitem[\protect\citeauthoryear{Wood}{2012}]{wood_stratocumulus_2012}
\begin{barticle}
\bauthor{\bsnm{Wood}, \binits{R.}}:
\batitle{Stratocumulus {Clouds}}.
\bjtitle{Monthly Weather Review}
\bvolume{140}(\bissue{8}),
\bfpage{2373}--\blpage{2423}
(\byear{2012})
\doiurl{10.1175/MWR-D-11-00121.1}
\end{barticle}
\endbibitem

\bibitem[\protect\citeauthoryear{Leon et~al.}{2008}]{leon_climatology_2008}
\begin{botherref}
\oauthor{\bsnm{Leon}, \binits{D.C.}},
\oauthor{\bsnm{Wang}, \binits{Z.}},
\oauthor{\bsnm{Liu}, \binits{D.}}:
Climatology of drizzle in marine boundary layer clouds based on 1 year of data from {CloudSat} and {Cloud}-{Aerosol} {Lidar} and {Infrared} {Pathfinder} {Satellite} {Observations} ({CALIPSO}).
Journal Of Geophysical Research-Atmospheres
\textbf{113}(D)
(2008)
\doiurl{10.1029/2008JD009835}
\end{botherref}
\endbibitem

\bibitem[\protect\citeauthoryear{Ackerman et~al.}{2003}]{ackerman_enhancement_2003}
\begin{barticle}
\bauthor{\bsnm{Ackerman}, \binits{A.S.}},
\bauthor{\bsnm{Toon}, \binits{O.B.}},
\bauthor{\bsnm{Stevens}, \binits{D.E.}},
\bauthor{\bsnm{Coakley}, \binits{J.A.}}:
\batitle{Enhancement of cloud cover and suppression of nocturnal drizzle in stratocumulus polluted by haze}.
\bjtitle{Geophysical Research Letters}
\bvolume{30}(\bissue{7}),
\bfpage{1381}
(\byear{2003})
\doiurl{10.1029/2002GL016634}
\end{barticle}
\endbibitem

\bibitem[\protect\citeauthoryear{Wang et~al.}{2011}]{wang_manipulating_2011}
\begin{barticle}
\bauthor{\bsnm{Wang}, \binits{H.}},
\bauthor{\bsnm{Rasch}, \binits{P.J.}},
\bauthor{\bsnm{Feingold}, \binits{G.}}:
\batitle{Manipulating marine stratocumulus cloud amount and albedo: a process-modelling study of aerosol-cloud-precipitation interactions in response to injection of cloud condensation nuclei}.
\bjtitle{Atmospheric Chemistry And Physics}
\bvolume{11}(\bissue{9}),
\bfpage{4237}--\blpage{4249}
(\byear{2011})
\doiurl{10.5194/acp-11-4237-2011}
\end{barticle}
\endbibitem

\bibitem[\protect\citeauthoryear{PINCUS and Baker}{1994}]{pincus_effect_1994}
\begin{barticle}
\bauthor{\bsnm{PINCUS}, \binits{R.}},
\bauthor{\bsnm{Baker}, \binits{M.B.}}:
\batitle{{EFFECT} {OF} {PRECIPITATION} {ON} {THE} {ALBEDO} {SUSCEPTIBILITY} {OF} {CLOUDS} {IN} {THE} {MARINE} {BOUNDARY}-{LAYER}}.
\bjtitle{Nature}
\bvolume{372}(\bissue{6503}),
\bfpage{250}--\blpage{252}
(\byear{1994})
\end{barticle}
\endbibitem

\bibitem[\protect\citeauthoryear{Christensen and Stephens}{2012}]{christensen_microphysical_2012}
\begin{botherref}
\oauthor{\bsnm{Christensen}, \binits{M.W.}},
\oauthor{\bsnm{Stephens}, \binits{G.L.}}:
Microphysical and macrophysical responses of marine stratocumulus polluted by underlying ships: 2. {Impacts} of haze on precipitating clouds.
Journal of Geophysical Research: Atmospheres
\textbf{117}(D11)
(2012)
\doiurl{10.1029/2011JD017125} .
\_eprint: https://onlinelibrary.wiley.com/doi/pdf/10.1029/2011JD017125.
Accessed 2024-10-08
\end{botherref}
\endbibitem

\bibitem[\protect\citeauthoryear{Yuan et~al.}{2011}]{yuan_microphysical_2011}
\begin{barticle}
\bauthor{\bsnm{Yuan}, \binits{T.}},
\bauthor{\bsnm{Remer}, \binits{L.A.}},
\bauthor{\bsnm{Yu}, \binits{H.}}:
\batitle{Microphysical, macrophysical and radiative signatures of volcanic aerosols in trade wind cumulus observed by the {A}-{Train}}.
\bjtitle{Atmospheric Chemistry And Physics}
\bvolume{11}(\bissue{14}),
\bfpage{7119}--\blpage{7132}
(\byear{2011})
\doiurl{10.5194/acp-11-7119-2011}
\end{barticle}
\endbibitem

\bibitem[\protect\citeauthoryear{Yuan et~al.}{2024}]{yuan_abrupt_2024}
\begin{barticle}
\bauthor{\bsnm{Yuan}, \binits{T.}},
\bauthor{\bsnm{Song}, \binits{H.}},
\bauthor{\bsnm{Oreopoulos}, \binits{L.}},
\bauthor{\bsnm{Wood}, \binits{R.}},
\bauthor{\bsnm{Bian}, \binits{H.}},
\bauthor{\bsnm{Breen}, \binits{K.}},
\bauthor{\bsnm{Chin}, \binits{M.}},
\bauthor{\bsnm{Yu}, \binits{H.}},
\bauthor{\bsnm{Barahona}, \binits{D.}},
\bauthor{\bsnm{Meyer}, \binits{K.}},
\bauthor{\bsnm{Platnick}, \binits{S.}}:
\batitle{Abrupt reduction in shipping emission as an inadvertent geoengineering termination shock produces substantial radiative warming}.
\bjtitle{Communications Earth \& Environment}
\bvolume{5}(\bissue{1}),
\bfpage{1}--\blpage{8}
(\byear{2024})
\doiurl{10.1038/s43247-024-01442-3} .
\bcomment{Publisher: Nature Publishing Group}.
Accessed 2024-09-18
\end{barticle}
\endbibitem

\bibitem[\protect\citeauthoryear{Kurach et~al.}{2018}]{kurach_large-scale_2018}
\begin{botherref}
\oauthor{\bsnm{Kurach}, \binits{K.}},
\oauthor{\bsnm{Lucic}, \binits{M.}},
\oauthor{\bsnm{Zhai}, \binits{X.}},
\oauthor{\bsnm{Michalski}, \binits{M.}},
\oauthor{\bsnm{Gelly}, \binits{S.}}:
A {Large}-{Scale} {Study} on {Regularization} and {Normalization} in {GANs}.
arXiv:1807.04720 [cs, stat]
(2018).
arXiv: 1807.04720.
Accessed 2019-05-22
\end{botherref}
\endbibitem

\bibitem[\protect\citeauthoryear{Goren and Rosenfeld}{2012}]{goren_satellite_2012}
\begin{botherref}
\oauthor{\bsnm{Goren}, \binits{T.}},
\oauthor{\bsnm{Rosenfeld}, \binits{D.}}:
Satellite observations of ship emission induced transitions from broken to closed cell marine stratocumulus over large areas.
Journal of Geophysical Research: Atmospheres
\textbf{117}(D17)
(2012)
\doiurl{10.1029/2012JD017981} .
\_eprint: https://onlinelibrary.wiley.com/doi/pdf/10.1029/2012JD017981.
Accessed 2022-01-15
\end{botherref}
\endbibitem

\bibitem[\protect\citeauthoryear{Smalley et~al.}{2024}]{smalley_diurnal_2024}
\begin{barticle}
\bauthor{\bsnm{Smalley}, \binits{K.M.}},
\bauthor{\bsnm{Lebsock}, \binits{M.D.}},
\bauthor{\bsnm{Eastman}, \binits{R.}}:
\batitle{Diurnal {Patterns} in the {Observed} {Cloud} {Liquid} {Water} {Path} {Response} to {Droplet} {Number} {Perturbations}}.
\bjtitle{GEOPHYSICAL RESEARCH LETTERS}
\bvolume{51}(\bissue{4}),
\bfpage{2023}--\blpage{107323}
(\byear{2024})
\doiurl{10.1029/2023GL107323} .
\bcomment{Num Pages: 9 Place: Washington Publisher: Amer Geophysical Union Web of Science ID: WOS:001159944100001}.
Accessed 2024-10-05
\end{barticle}
\endbibitem

\bibitem[\protect\citeauthoryear{Qiu et~al.}{2024}]{qiu_daytime_2024}
\begin{barticle}
\bauthor{\bsnm{Qiu}, \binits{S.}},
\bauthor{\bsnm{Zheng}, \binits{X.}},
\bauthor{\bsnm{Painemal}, \binits{D.}},
\bauthor{\bsnm{Terai}, \binits{C.R.}},
\bauthor{\bsnm{Zhou}, \binits{X.}}:
\batitle{Daytime variation in the aerosol indirect effect for warm marine boundary layer clouds in the eastern {North} {Atlantic}}.
\bjtitle{ATMOSPHERIC CHEMISTRY AND PHYSICS}
\bvolume{24}(\bissue{5}),
\bfpage{2913}--\blpage{2935}
(\byear{2024})
\doiurl{10.5194/acp-24-2913-2024} .
\bcomment{Num Pages: 23 Place: Gottingen Publisher: Copernicus Gesellschaft Mbh Web of Science ID: WOS:001190495200001}.
Accessed 2024-10-04
\end{barticle}
\endbibitem

\bibitem[\protect\citeauthoryear{Rosenfeld and Lensky}{1998}]{rosenfeld_satellite-based_1998}
\begin{barticle}
\bauthor{\bsnm{Rosenfeld}, \binits{D.}},
\bauthor{\bsnm{Lensky}, \binits{I.M.}}:
\batitle{Satellite-based insights into precipitation formation processes in continental and maritime convective clouds}.
\bjtitle{Bulletin Of The American Meteorological Society}
\bvolume{79}(\bissue{11}),
\bfpage{2457}--\blpage{2476}
(\byear{1998})
\end{barticle}
\endbibitem

\bibitem[\protect\citeauthoryear{Stein et~al.}{2015}]{stein_noaas_2015}
\begin{botherref}
\oauthor{\bsnm{Stein}, \binits{A.F.}},
\oauthor{\bsnm{Draxler}, \binits{R.R.}},
\oauthor{\bsnm{Rolph}, \binits{G.D.}},
\oauthor{\bsnm{Stunder}, \binits{B.J.B.}},
\oauthor{\bsnm{Cohen}, \binits{M.D.}},
\oauthor{\bsnm{Ngan}, \binits{F.}}:
{NOAA}’s {HYSPLIT} {Atmospheric} {Transport} and {Dispersion} {Modeling} {System}
(2015)
\doiurl{10.1175/BAMS-D-14-00110.1} .
Section: Bulletin of the American Meteorological Society.
Accessed 2024-10-11
\end{botherref}
\endbibitem

\bibitem[\protect\citeauthoryear{Gelaro et~al.}{2017}]{gelaro_modern-era_2017}
\begin{barticle}
\bauthor{\bsnm{Gelaro}, \binits{R.}},
\bauthor{\bsnm{McCarty}, \binits{W.}},
\bauthor{\bsnm{Suárez}, \binits{M.J.}},
\bauthor{\bsnm{Todling}, \binits{R.}},
\bauthor{\bsnm{Molod}, \binits{A.}},
\bauthor{\bsnm{Takacs}, \binits{L.}},
\bauthor{\bsnm{Randles}, \binits{C.A.}},
\bauthor{\bsnm{Darmenov}, \binits{A.}},
\bauthor{\bsnm{Bosilovich}, \binits{M.G.}},
\bauthor{\bsnm{Reichle}, \binits{R.}},
\bauthor{\bsnm{Wargan}, \binits{K.}},
\bauthor{\bsnm{Coy}, \binits{L.}},
\bauthor{\bsnm{Cullather}, \binits{R.}},
\bauthor{\bsnm{Draper}, \binits{C.}},
\bauthor{\bsnm{Akella}, \binits{S.}},
\bauthor{\bsnm{Buchard}, \binits{V.}},
\bauthor{\bsnm{Conaty}, \binits{A.}},
\bauthor{\bsnm{Silva}, \binits{A.M.}},
\bauthor{\bsnm{Gu}, \binits{W.}},
\bauthor{\bsnm{Kim}, \binits{G.-K.}},
\bauthor{\bsnm{Koster}, \binits{R.}},
\bauthor{\bsnm{Lucchesi}, \binits{R.}},
\bauthor{\bsnm{Merkova}, \binits{D.}},
\bauthor{\bsnm{Nielsen}, \binits{J.E.}},
\bauthor{\bsnm{Partyka}, \binits{G.}},
\bauthor{\bsnm{Pawson}, \binits{S.}},
\bauthor{\bsnm{Putman}, \binits{W.}},
\bauthor{\bsnm{Rienecker}, \binits{M.}},
\bauthor{\bsnm{Schubert}, \binits{S.D.}},
\bauthor{\bsnm{Sienkiewicz}, \binits{M.}},
\bauthor{\bsnm{Zhao}, \binits{B.}}:
\batitle{The {Modern}-{Era} {Retrospective} {Analysis} for {Research} and {Applications}, {Version} 2 ({MERRA}-2)}.
\bjtitle{Journal of Climate}
\bvolume{30}(\bissue{14}),
\bfpage{5419}--\blpage{5454}
(\byear{2017})
\doiurl{10.1175/JCLI-D-16-0758.1} .
Accessed 2019-10-17
\end{barticle}
\endbibitem

\bibitem[\protect\citeauthoryear{Platnick et~al.}{2017}]{platnick_modis_2017}
\begin{barticle}
\bauthor{\bsnm{Platnick}, \binits{S.}},
\bauthor{\bsnm{Meyer}, \binits{K.G.}},
\bauthor{\bsnm{King}, \binits{M.D.}},
\bauthor{\bsnm{Wind}, \binits{G.}},
\bauthor{\bsnm{Amarasinghe}, \binits{N.}},
\bauthor{\bsnm{Marchant}, \binits{B.}},
\bauthor{\bsnm{Arnold}, \binits{G.T.}},
\bauthor{\bsnm{Zhang}, \binits{Z.}},
\bauthor{\bsnm{Hubanks}, \binits{P.A.}},
\bauthor{\bsnm{Holz}, \binits{R.E.}},
\bauthor{\bsnm{Yang}, \binits{P.}},
\bauthor{\bsnm{Ridgway}, \binits{W.L.}},
\bauthor{\bsnm{Riedi}, \binits{J.}}:
\batitle{The {MODIS} {Cloud} {Optical} and {Microphysical} {Products}: {Collection} 6 {Updates} and {Examples} {From} {Terra} and {Aqua}}.
\bjtitle{IEEE Transactions on Geoscience and Remote Sensing}
\bvolume{55}(\bissue{1}),
\bfpage{502}--\blpage{525}
(\byear{2017})
\doiurl{10.1109/TGRS.2016.2610522}
\end{barticle}
\endbibitem

\bibitem[\protect\citeauthoryear{Grosvenor et~al.}{2018}]{grosvenor_remote_2018}
\begin{barticle}
\bauthor{\bsnm{Grosvenor}, \binits{D.P.}},
\bauthor{\bsnm{Sourdeval}, \binits{O.}},
\bauthor{\bsnm{Zuidema}, \binits{P.}},
\bauthor{\bsnm{Ackerman}, \binits{A.}},
\bauthor{\bsnm{Alexandrov}, \binits{M.D.}},
\bauthor{\bsnm{Bennartz}, \binits{R.}},
\bauthor{\bsnm{Boers}, \binits{R.}},
\bauthor{\bsnm{Cairns}, \binits{B.}},
\bauthor{\bsnm{Chiu}, \binits{J.C.}},
\bauthor{\bsnm{Christensen}, \binits{M.}},
\bauthor{\bsnm{Deneke}, \binits{H.}},
\bauthor{\bsnm{Diamond}, \binits{M.}},
\bauthor{\bsnm{Feingold}, \binits{G.}},
\bauthor{\bsnm{Fridlind}, \binits{A.}},
\bauthor{\bsnm{Hünerbein}, \binits{A.}},
\bauthor{\bsnm{Knist}, \binits{C.}},
\bauthor{\bsnm{Kollias}, \binits{P.}},
\bauthor{\bsnm{Marshak}, \binits{A.}},
\bauthor{\bsnm{McCoy}, \binits{D.}},
\bauthor{\bsnm{Merk}, \binits{D.}},
\bauthor{\bsnm{Painemal}, \binits{D.}},
\bauthor{\bsnm{Rausch}, \binits{J.}},
\bauthor{\bsnm{Rosenfeld}, \binits{D.}},
\bauthor{\bsnm{Russchenberg}, \binits{H.}},
\bauthor{\bsnm{Seifert}, \binits{P.}},
\bauthor{\bsnm{Sinclair}, \binits{K.}},
\bauthor{\bsnm{Stier}, \binits{P.}},
\bauthor{\bsnm{van Diedenhoven}, \binits{B.}},
\bauthor{\bsnm{Wendisch}, \binits{M.}},
\bauthor{\bsnm{Werner}, \binits{F.}},
\bauthor{\bsnm{Wood}, \binits{R.}},
\bauthor{\bsnm{Zhang}, \binits{Z.}},
\bauthor{\bsnm{Quaas}, \binits{J.}}:
\batitle{Remote {Sensing} of {Droplet} {Number} {Concentration} in {Warm} {Clouds}: {A} {Review} of the {Current} {State} of {Knowledge} and {Perspectives}}.
\bjtitle{Reviews of Geophysics}
\bvolume{56}(\bissue{2}),
\bfpage{409}--\blpage{453}
(\byear{2018})
\doiurl{10.1029/2017RG000593} .
Accessed 2021-08-25
\end{barticle}
\endbibitem

\end{thebibliography}

\end{document}